\documentclass[10pt, aps, prb, twocolumn, superscriptaddress, longbibliography]{revtex4-2}

\usepackage{xcolor}
\usepackage{graphicx}
\usepackage{amsfonts}
\usepackage{amssymb}
\usepackage{amsmath}
\usepackage{csquotes}
\usepackage{mathtools}
\usepackage{braket}
\usepackage{hyphenat}
\usepackage{siunitx}
\usepackage{microtype}

\usepackage{pdfpages}
\makeatletter
\AtBeginDocument{\let\LS@rot\@undefined}
\makeatother

\makeatletter
\AddToHook{cmd/appendix/before}{\def\cref@section@alias{appendix}\def\cref@subsection@alias{appendix}}
\makeatother

\usepackage{tabularray}
\usepackage[acronym]{glossaries}

\usepackage[english]{babel}
\usepackage{microtype}
\usepackage{hyperref}
\usepackage[english,capitalise]{cleveref}

\Crefname{appendix}{App.}{Apps.}
\Crefname{section}{Sec.}{Secs.}

\usepackage{orcidlink}
\usepackage{lipsum}

\newcommand{\refcite}[1]{Ref.~\cite{#1}}

\DeclareMathOperator{\tr}{tr}
\DeclareMathOperator{\hc}{h.c.}
\newcommand{\transpose}{{\mkern-1.5mu\text{T}}}
\newcommand{\abs}[1]{\vert #1 \vert}
\renewcommand\vec\boldsymbol
\renewcommand\Re{\mathrm{Re}\,}
\newcommand\ketbra[1]{\ket{#1}\!\bra{#1}}
\newcommand{\updownarrows}{{\uparrow\mathrel{\mspace{-3.5mu}}\downarrow}}
\DeclarePairedDelimiterX{\norm}[1]{\lVert}{\rVert}{#1}
\newcommand\idmatrix{1}
\newcommand\dd{\mathrm{d}}
\newcommand\vect[1]{\vert #1 )}

\DeclareMathOperator*{\argmin}{arg\,min}
\newcommand\package\textsc

\begin{document}

\newcommand\tudresdenaffil{Institute of Theoretical Physics, Technische Universität Dresden, 01062 Dresden, Germany}
\newcommand\unikoelnaffil{Institut für Theoretische Physik, Universität zu Köln, 50937 Cologne, Germany}
\newcommand\ctqmataffil{Würzburg-Dresden Cluster of Excellence ct.qmat, 01062 Dresden, Germany}
\newcommand\mpipksaffil{Max Planck Institute for the Physics of Complex Systems, Nöthnitzer Str.~38, 01187 Dresden, Germany}

\title{\textls[-15]{Diffusion in quantum state preparation: From passive cooling to system-bath engineering}}

\author{Tim Pokart\,\orcidlink{0009-0006-1415-3261}}
\affiliation{\tudresdenaffil}
\author{Lukas König}
\affiliation{\tudresdenaffil}
\author{Sebastian Diehl\,\orcidlink{0009-0009-5631-9882}}
\affiliation{\unikoelnaffil}
\email{diehl@thp.uni-koeln.de}
\author{Jan Carl Budich\,\orcidlink{0000-0002-9859-9626}}
\affiliation{\tudresdenaffil}
\affiliation{\mpipksaffil}
\affiliation{\ctqmataffil}

\begin{abstract}
    We investigate and compare two particle number conserving protocols for the preparation of a topologically nontrivial state. The first is derived from thermally coupling the system to a cold bath, while the second is based on engineered dissipation. We numerically study the time required to reach the target state 
    as well as its robustness against physically important perturbations. Crucially, in both protocols the cooling capability is limited by dissipatively induced diffusion processes. The resulting quadratic scaling of the cooling time with system size is also corroborated analytically using mean-field approximations and a purely classical random-walk model. Furthermore, we find that the engineered protocol admits a unique and stable dark state, which contributes to an ongoing discussion regarding the applicability of dissipative state preparation to many-body systems.
\end{abstract}

\hyphenation{Go-ri-ni Kos-sa-kow-ski Su-dar-shan Lind-blad}
\newacronym{gksl}{GKSL}{Gorini\allowbreak{}--\allowbreak{}Kossakowski\allowbreak{}--\allowbreak{}Sudarshan\allowbreak{}--\allowbreak{}Lindblad}
\newacronym{ssh}{SSH}{Su--Schrieffer--Heeger}
\newacronym{dmrg}{DMRG}{density matrix renormalization group}
\newacronym{ed}{ED}{exact diagonalization}
\newacronym{ule}{ULE}{universal Lindblad equation}
\newacronym{fkpp}{FKPP}{Fisher\allowbreak{}--\allowbreak{}Kolmogorov\allowbreak{}--\allowbreak{}Petrovsky\allowbreak{}--\allowbreak{}Piskunov}
\newacronym{kms}{KMS}{Kubo\allowbreak{}--Martin\allowbreak{}--Schwinger}
\maketitle

\date{\today}
\section{Introduction}

Robust and efficient protocols for quantum state preparation are of key importance in quantum science. 
Applications range from universal quantum information processing paradigms~\cite{Feynman1982,doi:10.1126/science.273.5278.1073,Ladd2010,Kim2023,RevModPhys.90.015002,10.1145/1219092.1219096,DiVincenzo2000} to the simulation of low-temperature physics in complex quantum many-body systems~\cite{Cirac2012,Daley2022}. 
In particular, preparing the ground state of a generic local Hamiltonian already contains unitary circuit based quantum computing as special case~\cite{doi:10.1137/S0097539704445226,Aharonov2009,preskill2023quantum,Watrous2009}.
Turning to quantum simulation~\cite{doi:10.1126/science.1177838,RevModPhys.86.153,PRXQuantum.2.017003}, the complexity of state preparation currently entails that experimental implementation of many-body Hamiltonians~\cite{Gross2017,Guo2021,PhysRevX.8.011002,Dreon2022,Kongkhambut2022,Landig2016,PhysRevLett.115.230403,PhysRevLett.131.243401,PhysRevX.15.021089,Periwal2021,Young2024,Sauerwein2023} is by far more developed than the preparation of their low-temperature states~\cite{PRXQuantum.2.017003,Houck2012}, despite impressive progress on the latter~\cite{Mazurenko2017,PhysRevLett.99.200405,Schindewolf2022}. %

Dissipation is crucial for preparing low-entropy quantum states.
For conventional cooling, e.g., via cryogenic fridges~\cite{Pagano_2019,PhysRevLett.125.260502} or lasers~\cite{RevModPhys.93.025001,Gaerttner2017}, it is paramount for relaxation toward low energy. 
However, in quantum simulators realized with synthetic quantum matter, this approach tends to be more severely limited than in conventional electronic materials~\cite{CHANG2023100054}. 
In addition, the hardness of state preparation is reflected in fundamental ergodicity bottlenecks manifesting in divergent dwelling times of dissipative dynamics in energetic local minima~\cite{Chen2025}.

\begin{figure}[t]
    \centering
    \includegraphics{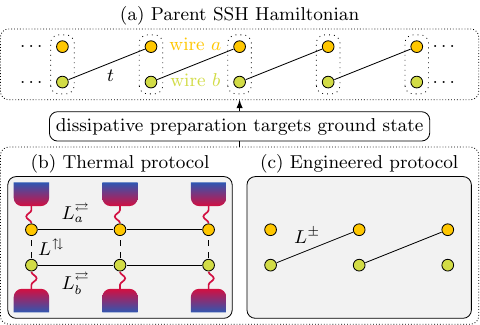}
    \caption{(a) Parent Hamiltonian $H$ from \cref{eq:common_hamiltonian} with yellow $a$ (green $b$) sub-lattice sites, and inter-cell hopping $t$. (b) Generic cooling protocol based on local system-bath couplings with a system governed by the parent Hamiltonian and a cold thermal environment inducing intra-cell (inter-cell) cooling $L^\updownarrows$ ($L^{\rightleftarrows}_{a,b}$). (c) Engineered dissipative state preparation protocol targeting the ground state of a specific parent Hamiltonian in a non-equilibrium fashion, resulting in Lindblad jump operators $L^\pm$ that pump toward the steady state without relying on coherent Hamiltonian dynamics.}\label{fig:cooling-protocols}
\end{figure}

With \emph{engineered dissipation}, a different paradigm for quantum cooling protocols has been developed, which harnesses dissipation as an active resource~\cite{Diehl2008,Verstraete2009,PhysRevA.78.042307,Mueller2012,Harrington2022}:
Instead of merely inducing decoherence by interactions with an uncontrollable environment~\cite{Breuer2007, Daley2014,Duan1998,DiVincenzo1998,Gyongyosi2019}, dissipation may be utilized to remove entropy and thus cool largely arbitrary initial states toward pure quantum states of matter~\cite{PhysRevA.78.042307,Mueller2012,Harrington2022,Daffer2003,Zanardi1998, Pichler2010, Griessner2006,Griessner2007,Albash2015} (see \cref{fig:cooling-protocols}).
This approach is particularly suitable for highly controllable synthetic quantum matter, where an impressive experimental toolbox affords carefully designed system-environment couplings that realize the desired non-equilibrium dynamics~\cite{PhysRevLett.77.4728}.
While the focus of this work is on \emph{dissipative state preparation} protocols in which the structure of the desired state is explicitly embedded in the generator of the dynamics, we note that less specialized protocols with a priori unknown target have also been developed~\cite{Molpeceres2025,PRXQuantum.6.010361,science.adh9932}.
The experimental applicability of this approach for many-body systems is closely related to the asymptotic scaling of preparation-times with system size as well as the robustness of the targeted steady state with respect to physically relevant perturbations.

In this article, we compare two \emph{particle number-conserving} dissipative protocols for preparing the ground state of a generalized \gls{ssh} model (see illustration in \cref{fig:cooling-protocols}) as a paradigmatic example of topological insulators~\cite{PhysRevLett.124.240404,PhysRevResearch.3.043119,PhysRevB.107.174312,Lyublinskaya_2023,Lyublinskaya_2025,baburin2025minimalvelocitytravellingwave} and topological superconductors~\cite{shustin2025dissipation}.
The first protocol mimics a cryostat, where the coupled dynamics aim at equilibrating the system with a low-temperature heat bath.
This provides a generic formulation of cooling dynamics within the framework of local thermal perturbations~\cite{Chen2025}, cf.~\cref{fig:cooling-protocols}\,(b). 
The second approach exemplifies engineered dissipation with a set of experimentally accessible channels for dissipative state preparation, where the \gls{ssh} ground state is reached as the pure steady state (dark state) of a quantum master equation, cf.~\cref{fig:cooling-protocols}\,(c). 

In brief, we find---for both approaches---a quadratic scaling of the cooling time $\tau$ in system size $N$, i.e., $\tau \sim N^2$. 
This is in sharp contrast to Gaussian dissipative models that predict a constant $\tau$~\cite{Diehl2008}.
The origin of this different behavior is particle number conservation, which is absent in the Gaussian model, but present here.
Specifically, the quadratic scaling is compatible with the physical picture that long-time dynamics is dominated by diffusion in our number-conserving setting (see Refs.~\cite{PhysRevB.107.174312,Lyublinskaya_2023} for the engineered dissipation case). 
We underpin this intuition comparing extensive numerical studies to both a number-conserving mean-field description and a classical random-walk model, which are shown to capture the non-equilibrium dynamics quite accurately.
In addition, we discuss the temperature dependence of the thermal protocol, and show the uniqueness as well as robustness of the engineered steady state.
Quantitatively, we argue that the engineered protocol may reach the ground state much faster than the thermal protocol as measured by the diffusion constant $D$ in $\tau = D N^2$.

The rest of the article is structured as follows: In \cref{sec:model} the target Hamiltonian as well as both thermal and engineered dissipation protocols are introduced.
For both protocols, we perform extensive numerical calculations, with methods discussed in \cref{sec:methods} and their results summarized in \cref{sec:results}.
In \cref{sec:diffusion}, we link the observed scaling of the cooling time $\tau$ to the diffusive nature of the cooling physics by introducing and validating the aforementioned mean-field description as well as a random-walk model, thus explaining the quadratic scaling.
We conclude the paper with a comparison to recent analytical work in \cref{sec:comparison} and subsequent discussion in \cref{sec:discussion}.

\section{System and Dissipative Protocols}\label{sec:model}

We are interested in the system illustrated in \cref{fig:cooling-protocols}, which is composed of fermions on a lattice given by two wires with $L$ sites each, such that the total system contains $N=2L$ sites, subject to periodic boundary conditions and nearest-neighbor interaction.
We target the ground state of the dimerized \gls{ssh} Hamiltonian
\begin{align}
    H=\sum_{j=1}^L t c^\dagger_{j+1,a} c_{j,b} +\hc \label{eq:common_hamiltonian}
\end{align}
where $c_{j,\gamma}$ annihilate a fermion at site $j$ of the wire $\gamma=a,b$; the inter cell hopping is given by the coefficient $t = 1$.
The system is at half filling, such that there are $L$ fermions in total on the lattice; all dynamics are fermion number conserving.
To diagonalize the dimerized \gls{ssh} Hamiltonian, it is convenient to introduce the (anti)symmetric bond operators
\begin{align}
    x_{j,\pm} = \frac{1}{\sqrt{2}} (c_{j,b} \pm c_{j+1,a}), 
    \label{eq:bondbasis}
\end{align}
such that $H = \sum_{j} t (x_{j,+}^\dagger x_{j,+} - x_{j,-}^\dagger x_{j,-})$.
The $x_{j,\pm}$ operators are themselves fermionic, i.e., they anticommute $\{x^\dagger_{\sigma, j}, x_{\sigma', j'}\} = \delta_{\sigma \sigma'} \delta_{j j'}$, and thus the total fermion number $\mathcal{N} = \sum_j n_j$ can then be split into $\mathcal{N} = \mathcal{N}_+ + \mathcal{N}_-$ with
\begin{align}
    \mathcal{N}_\pm = \sum_j x_{j,\pm}^\dagger x_{j,\pm}.
\end{align}

The dissipative dynamics in an open quantum system are governed by the \gls{gksl} equation~\cite{Lindblad1976,GoriniKossakowskiSudarshan1976}
\begin{align}
    \dot\rho = \mathcal{L}[\rho] = -i [H,\rho] + \sum_\mu \Big(L_\mu \rho L_\mu^\dagger - \frac{1}{2} \{L_\mu^\dagger L_\mu, \rho\}\Big).\label{eq:Lindblad}
\end{align}
Different choices for the jump operators $L_\mu$ generate different cooling dynamics.
In our case one choice will lead to dynamics corresponding to a thermal coupling and the other corresponds to an engineered system-bath interaction.
How fast the dissipative dynamics reach the ground state, i.e., the cooling rate of the protocol, is dictated by the Liouvillian gap.
Let $\lambda_i$ be the eigenvalues of the Liouvillian $\mathcal L$ sorted in decreasing order by their real part ($\Re \lambda_0 > \Re \lambda_1 \geq \ldots$); then $\lambda_0 = 0$ is the steady state and we define the Liouvillian gap $\Delta \equiv \abs{\Re \lambda_1}$.
As the mode that takes the longest to decay into the steady state has a coefficient $\propto \exp(-\lambda_1 t)$, the equilibration time scales are given by the characteristic cooling time 
\begin{align}
    \tau \equiv \Delta^{-1} = \frac{1}{\abs{\Re \lambda_1}}.
    \label{eqn:taudelta}
\end{align}
We introduce the Lindblad operators describing the thermal dissipation dynamics in \cref{sec:model:thermal}, while \cref{sec:model:engineered} concerns the engineered dissipation case.

\subsection{Thermal Dissipation}\label{sec:model:thermal}

When describing open quantum system dynamics caused by thermally coupling the system to a Markovian bath, several different approaches exist that bring the Redfield equation, i.e., the quantum master equation involving only the Born-Markov approximation, into \gls{gksl} form~\cite{Shiraishi2024}.
Their fundamental differences lie in the specific approximations used to ensure complete positivity when deriving the Lindblad operators $L_\mu$ from an interaction Hamiltonian of the form $H_\text{int}=\sum_\mu X_\mu\otimes B_\mu$ with $X_\mu$ ($B_\mu$) an operator on the system (bath).
To provide a broad comparison, we implement both the Davies equation and the \gls{ule}~\cite{Davies1974,Nathan2020}.
The latter generates one jump operator per cooling channel, whereas the former generates a finer decomposition by additionally assigning separate operators to each individual internal energy transition of the system.

\paragraph*{Modelling the bath.}
In order to obtain the Lindblad operators, we need to make assumptions about the physics involved in the interaction with the bath.
To this end, we assume that in the weak-coupling limit we are concerned with, the baths are sufficiently local such that each site is effectively coupled to its own bath as illustrated in \cref{fig:cooling-protocols}.
Although the exact choice of dissipators to model the weak thermal coupling changes the quantitative behavior of the theory, the results are qualitatively unchanged, as long as the system-bath interaction is (i) sufficiently local, (ii) the bath relaxes much faster than the system but (iii) interactions happen on similar energy scales as the system itself has.

A platform capable of cooling toward the desired state requires a system-bath interaction that preserves the total particle number and enables both local particle number equilibration and global thermalization. 
Physically, this may be implemented by coupling to ancillary thermal degrees of freedom such as a phononic bath or auxiliary sites. 
For concreteness, we model the effective behavior of such couplings with the following allowed nearest-neighbor site\hyp{}switching operators:
\begin{align}
    X^{\rightleftarrows}_{j,\gamma}&=c_{j,\gamma}^\dagger c_{j\pm 1,\gamma}, &
    X^\updownarrows_{j,\gamma}&=c_{j,\gamma}^\dagger c_{j,1-\gamma}, \label{eq:thermal_system_operators}
\end{align}
where $X^\rightarrow$ ($X^\leftarrow$) mediates hopping to the right (left) along the wires and $X^\updownarrows$ swaps particles in one unit cell; in total, there are $6L$ such operators. 
As the baths themselves are assumed to be separate, all off-diagonal correlation functions effectively vanish~\footnote{In the case of the Davies equation, this will render the power spectrum diagonal as $\gamma_{\mu\nu}(\omega) \propto \delta_{\mu\nu}$. For the \gls{ule}, we then assume a diagonal spectral density function $J_{\mu\nu}\propto \delta_{\mu\nu}$.}.

\paragraph*{Davies equation.}
To arrive at the Davies equation~\cite{Davies1974}, one additionally assumes that the rotating wave approximation holds, i.e., all energy scales of the system given are much smaller than those of the bath.
The Lindblad operators describing the system are then~\cite{Breuer2007,Shiraishi2024}
\begin{align}
    L_\mu(\omega) = \sqrt{J(\omega)} \sum_{\mathclap{\epsilon_n-\epsilon_m=\omega}}\ket{m}\braket{m|X_\mu|n}\bra{n}\label{eq:Adecomp},
\end{align}
where the $\ket{n}$, $\ket{m}$ are eigenstates to the eigenenergies $\epsilon_n$, $\epsilon_m$ of the system's Hamiltonian in \cref{eq:common_hamiltonian} and $J(\omega)$ is the spectral density function.  
The flat-band nature of the model prevents typical pitfalls when applying the Davies equation to many-body systems; namely, that the gap generically closing with increasing system size renders the rotating wave approximation invalid and that typically numerical studies are limited by the cost of the exponentially many jump operators.

\begin{figure}
    \centering
    \includegraphics{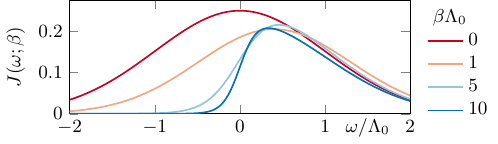}
    \caption{Plot of the spectral density $J(\omega; \beta)$ in \cref{eq:spectral_density} for different values of $\beta$ in units of the typical energy scales of the model $\Lambda_0$. For small temperatures $T \lesssim \Lambda_0/5$ ($\beta\Lambda_0\gtrsim 5$), for which the ground state constitutes a macroscopic portion of the steady state, the contributions increasing energy at $\omega = -\Lambda_0$ are heavily suppressed.}
    \label{fig:spectral-density}
\end{figure}
\paragraph*{Universal Lindblad equation.}
Recently, multiple different approaches have been found that microscopically derive the Lindblad equation while avoiding the rotating wave approximation~\cite{Nathan2020,Mozgunov2019,Davidovic2020}. 
We choose the \gls{ule}~\cite{Nathan2020} modeling the action of the bath by prescribing only one Lindblad operator 
\begin{equation}
    L_\mu=\sum_{n,m}\sqrt{J(\epsilon_n-\epsilon_m)}\ket{m}\braket{m|X_\mu|n}\bra{n}\label{eq:ULE}
\end{equation}
for each system-bath interaction, which incoherently overlays the different energy transition contributions, again weighted by the spectral density function $J(\omega)$.

\paragraph*{Spectral density function.}
The spectral density function in \cref{eq:Adecomp,eq:ULE} is chosen such that the transition weights produce Glauber dynamics~\cite{Chen2025}, with
\begin{align}
    J(\omega; \beta)=\frac{1}{2+\ln(1+\beta\Lambda_0)}\frac{e^{-\omega^2/(2\Lambda_0^2)}}{1+e^{-\beta\omega}}. \label{eq:spectral_density} 
\end{align} 
Here, $\beta$ is an inverse temperature and $\Lambda_0=2t$ is a relevant energy scale of the theory, which we set to the level spacing.
The spectral density is plotted for different values of $\beta$ in \cref{fig:spectral-density}.
Because the spectral density functions are derived from the correlation functions of the bath operators, they must fulfill the \gls{kms} condition~\cite{Mozgunov2019}
\begin{align}
    J(\omega; \beta)/J(-\omega;\beta)=e^{\beta\omega}.\label{eq:kubomartinschwinger}
\end{align}
All functions fulfilling this condition and the normalization $0<J(\omega)<1$ should lead to broadly comparable results~\cite{Chen2025}, although they imply different properties of the baths.


\begin{table*}
    \begin{tblr}{
        colspec={l l X},
        row{1} = {font=\bfseries}, 
        row{6} = {font=\bfseries}, 
    }
        \hline
        Numerical Method &  & Description \\ \hline
        \hyperref[sec:methods:vectorization]{Exact Vectorization}  & $N \leq 14$ & Access to the full spectrum and exact dynamics; is numerically robust and can rule out additional stationary solutions. \\
        \hyperref[sec:methods:numerical]{Quantum trajectories} & $N \leq 24$ & Shows potential instability. Results are statistical with sampling uncertainty.  \\
        \hyperref[sec:methods:ths]{Truncated Hilbert space} & $N \leq 40$  & Lower bound on cooling times; also helpful for analytical constructions. \\
        \hyperref[sec:methods:vectorization:tn]{Tensor networks} & $N \leq 60$ & Unbiased. Lack of variational principle requires other methods for verification. \\
        \hline\hline
        Analytical Method &  & Description \\ \hline
        \hyperref[sec:diffusion:mean-field]{Mean-field theory} & Continuous & Prediction of diffusion parameters without any instability. \\
        \hyperref[sec:diffusion:random-walk]{Random-walk model} & Classical & Intuitive picture of the dynamics. \\ \hline
    \end{tblr}
    \caption{Overview of the various numerical and analytical methods employed in this work. For the numerical methods, typical systems sizes are indicated. For tensor networks, the verification is performed by comparing to the previous exact methods and lower bounds as well as spectral gradient search. The mean-field theory is intrinsically continuous while the random-walk model provides a purely classical description on a discrete lattice.}\label{tab:overview}
\end{table*}

\subsection{Engineered Dissipation}\label{sec:model:engineered}
As an experimentally tractable model~\cite{PhysRevLett.77.4728,Awschalom2013,Harrington2022}, we study a scheme of explicitly local engineered operators on the two-legged ladder. 
The key difference to the thermal dissipation introduced previously is that the system's internal dynamics governed by the Hamiltonian can be realized to be much smaller than the dynamics of the engineered dissipation, thus leading to a large separation between their time scales.
Therefore, these systems are effectively frozen such that $H=0$.
Following \refcite{Altland2021}, the engineered cooling protocol is then given by the two kinds of Lindblad jump operators 
\begin{align}
    L^+_{j,\gamma} &= \sqrt{\kappa} c_{j, \gamma}^\dagger x_{j,+}, &
    L^-_{j,\gamma} &= \sqrt{\kappa} c_{j, \gamma} x^\dagger_{j,-}.
    \label{eq:engineered_jump_operators}
\end{align}
In total, there are $4L$ jump operators, which all annihilate the dark state
\begin{align*}
    \ket{\circ} = \prod_j x_{j,-}^\dagger \ket{0},
\end{align*}
such that the steady state is given by $\rho_0 = \ketbra{\circ}$.
Intuitively, the jump operators may be interpreted as two components of a pump from the excited to the ground states and the steady state is reached as follows: By the action of $L^+$ any contribution in the coherent $x_+$ mode will be depleted and instead a single incoherent fermion will be placed in the subwires.
As a second step, the $L^-$ turn these fermions into coherent $x_-$ states. 

\paragraph*{A note on uniqueness.}
Recently, there has been debate~\cite{PhysRevB.107.174312,Lyublinskaya_2023,Lyublinskaya_2025,baburin2025minimalvelocitytravellingwave} on whether the family of engineered protocols to which the one in \cref{eq:engineered_jump_operators} belongs is stable, i.e., whether there exist additional stationary solutions to the Lindblad equation, which could compete with the dark state.
For example, such an instability scenario occurs when the jump operators are restricted to $L^+_a$ only~\cite{PhysRevB.107.174312}, where
\begin{align*}
    \ket{\square} = \sum_n (-1)^n \sum_{\abs{\vec k}=n} (x_{k_1,+}^\dagger x_{k_1,-}) \cdots (x_{k_n,+}^\dagger x_{k_n,-}) \ket{\circ} 
\end{align*}
emerges as a second dark state, see \cref{app:uniqueness:second_dark_state} for a proof.
However, already a sufficiently generic subset of the Lindblad operators admits the construction of a jump sequence connecting every state in Hilbert space to the dark state $\ket{\circ}$.
Based on Theorem 2 in Ref.~\cite{PhysRevA.78.042307}, the existence of such sequences proves that $\ket{\circ}$ is the unique steady state, thus ruling out other (pure or mixed) stationary solutions.
Intuitively, this is due to these jump sequences necessarily occurring in the Taylor expansion of $\exp(\mathcal{L} t)$ for finite times $t$, thus slowly draining any non-dark state amplitudes in the state; for details we refer to \cref{app:uniqueness}.

\section{Methodology}\label{sec:methods}

For the two cooling protocols introduced above, we aim to determine (i) whether they converge to a unique steady state and, if so, (ii) the rate at which this steady state is approached.
For this, we rely on both analytical arguments presented in the next section as well as several numerical methods.
In the following, we give a short overview of the latter, which is summarized in \cref{tab:overview}.
Concretely, we employ \gls{ed} and \gls{dmrg} as outlined in \cref{sec:methods:vectorization} to study the full spectrum.
This is complemented by quantum trajectories in \cref{sec:methods:numerical} and the construction of a truncated Hilbert space approach in \cref{sec:methods:ths}.

\subsection{Vectorization}\label{sec:methods:vectorization}

The most straightforward way to solve the \gls{gksl} equation is by vectorizing the density matrix using the isomorphism $\rho = \sum \rho_{ij} \ket{i}\!\bra{j} \rightarrow \vect\rho = \sum \rho_{ij} \ket{i}\!\ket{j}$. 
The superoperator $\mathcal L$ then appears in the form of a matrix 
\begin{align*}
    \tilde {\mathcal L} &= -i (\idmatrix \otimes H - H^\transpose\otimes\idmatrix) \nonumber\\
    &\qquad+ \sum_j (L_j^\ast \otimes L_j - \frac{1}{2} (\idmatrix \otimes L_j^\dagger L_j + L_j^\transpose L_j^\ast \otimes \idmatrix)).
\end{align*}
With this construction, \cref{eq:Lindblad} turns into $\dd_t \vect \rho = \tilde{\mathcal L} \vect \rho$ and the eigenvalues of the Liouvillian can be accessed by diagonalization of $\tilde{\mathcal L}$.
However, this approach incurs the critical drawback of squaring the Hilbert space dimension involved in the computation.
This restricts computations involving \gls{ed} to systems of size $N \approxeq 14$.

\subsubsection{Tensor Network Methods}\label{sec:methods:vectorization:tn}
As the system itself is one-dimensional and also sufficiently local, for larger systems, we employ \gls{dmrg} to obtain the cooling time $\tau$.
Although \gls{dmrg} is a well-established method for Hermitian systems, the Lindbladian is non-Hermitian, rendering \gls{dmrg} ill behaved.
Several tensor-network approaches exist that are in principle applicable to the Liouvillian~\cite{guo2022variationalmatrixproductstate,5vnl-w9p4,PhysRevB.105.205125,PhysRevLett.130.100401,westhoff2025tensornetworkframeworklindbladian,PhysRevResearch.6.043182,sander2025largescalestochasticsimulationopen}; however, we found them to be less reliable for our particular system than optimizing the Rayleigh quotient $(\rho|\tilde{\mathcal{L}}|\rho)$ using \gls{dmrg} with a local eigensolver capable of solving non-Hermitian problems, i.e., an Arnoldi solver~\cite{Causer2025,PhysRevA.92.022116}. 
While not rigorously founded, the results are stable and in good agreement with data obtained from the complementary methods.
To obtain a vector with large overlap in the eigenspace associated with $\lambda_1$, we amend the Liouvillian by a term punishing the overlap with the steady state $\rho_0$, i.e., adjusting $\tilde{\mathcal L} \rightarrow \tilde{\mathcal L} - w |\rho_0)(\rho_0|$ for sufficiently large $w = \mathcal O(10\lambda_1)$.
The representative found in this way is neither orthogonal to $|\rho_0)$ nor the exact eigenvector.
Extracting the Liouvillian gap can be achieved by computing the effective Liouvillian $\hat{\mathcal{L}}_{ij} = (\rho_i|\tilde{\mathcal{L}}|\rho_j)$ as well as the Gram matrix $G_{ij} = (\rho_i|\rho_j)$ and solving the generalized eigenvalue problem $(\hat{\mathcal{L}}, G)$.
Equivalently, a Löwdin (or symmetric) orthogonalization~\cite{LOWDIN1970185} on $G$ may be performed to form an orthonormal basis.

\subsubsection{Spectral Gradient Search}
We may check the results obtained from \gls{dmrg} by comparing with \gls{ed} results for small systems and the quantum trajectories for intermediate systems; for both, we find perfect agreement in the respective overlapping regimes.
Nevertheless, we wish to extend beyond this, as there are intrinsic sampling errors to quantum trajectories, the verification is limited to system sizes $N \approxeq 24$, and the modification $w |\rho_0)(\rho_0|$ may accidentally change $\lambda_1$ significantly as the penalty term is not tilted, i.e., the correct eigenspace would be targeted by $|\rho_0)(\idmatrix|$, which is, however, less stable in practice. 

\begin{figure}
    \centering
    \includegraphics{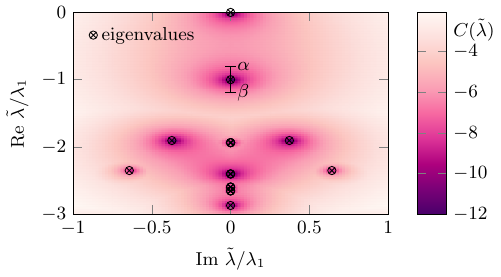}
    \caption{The cost function $C(\tilde \lambda)$ for the engineered dissipation Liouvillian with $N=8$. The eigenvalues are marked by stars, which coincide perfectly with the minima of the cost function. Crucially, this landscape is free of anomalous minima which motivates the spectral gradient search strategy. There, we start with the trust region $(\alpha, \beta)$ and successively shrink the interval using binary search.}\label{fig:liouvillian_spectrum_costfunction}
\end{figure}

To remedy these shortcomings, we introduce the following \emph{spectral gradient search} procedure.
Given a guess $\tilde \lambda$ for the eigenvalue of the Liouvillian $\tilde {\mathcal L}$, consider the cost function 
\begin{align*}
    C(\tilde \lambda) = \min_{|x)} \ln\ \norm{(\tilde{\mathcal L} - \tilde \lambda) |x)}^2,
\end{align*} 
which approaches $C(\tilde \lambda) \rightarrow -\infty$ for $\tilde \lambda$ converging to an actual eigenvalue of $\tilde{\mathcal L}$ as illustrated in \cref{fig:liouvillian_spectrum_costfunction}.
Now, the minima of $C(\tilde \lambda)$ can be obtained by solving for the roots of its gradient
\begin{align*}
    \partial_{\tilde{\lambda}^\ast}  C(\tilde \lambda) = -\frac{(x|\tilde{\mathcal L} - \tilde \lambda|x)}{\norm{(\tilde{\mathcal L} - \tilde \lambda) |x)}^2},
\end{align*}
given $|x)$ that minimizes the residual.
The minimization of the residual is itself a Hermitian minimization problem of the operator 
\begin{align}
    X_{\tilde\lambda} = (\tilde{\mathcal L}^\dagger - \tilde{\lambda}^\ast) (\tilde{\mathcal L} - \tilde \lambda), \label{eq:sgs_xlambda}
\end{align}
which is well-defined for tensor-network approaches such as \gls{dmrg}.
To find the root of the gradient, we utilize the fact that it is known that $\lambda_1$ is on the real line, such that starting from an initial guess within the trust region $(\alpha, \beta)$, we perform binary search, thereby shrinking the interval until the minimization routine fails.
Improving convergence speed of this method can be achieved by the preconditioning outlined in \cref{app:sgs_preconditioning}.
We note that gradient descent on a function similar to $C(\tilde \lambda)$ has been proposed for obtaining the spectrum of non-Hermitian operators~\cite{guo2022variationalmatrixproductstate}, but is slower and less reliable in practice for our system. 

\subsection{Quantum Trajectories}\label{sec:methods:numerical}

For intermediate systems we provide data from the quantum jump method, which will also serve as a fruitful framework to understand the mechanism producing quadratic gap scaling observed. 
Underlying this approach is the quantum trajectory picture that will be briefly introduced in the following.
The key insight is that the deterministic \gls{gksl} \cref{eq:Lindblad} governing the dynamics of a density matrix can equivalently be described by a stochastic equation of the system state~\cite{Breuer2007}. A measurement of the density matrix can then be replicated by averaging the measurements over many realizations of that stochastic process. 
In this picture, the full density matrix only describes information of the quantum state ensemble involved in the process.
By rewriting the \gls{gksl} equation as 
\begin{align*}
    \mathcal{L}[\rho] = -i (\tilde{H} \rho - \rho \tilde{H}^\dagger) + \sum_\mu L_\mu \rho L_\mu^\dagger,
\end{align*}
with $\tilde H = H - \frac{i}{2} \sum_\mu L_\mu^\dagger L_\mu$ the non-Hermitian effective Hamiltonian, the full dynamics are captured by an unbiased sampling of initial states ($\ket{\phi(0)}$) and their subsequent time evolution ($\ket{\phi(t)}$) with respect to $\tilde H$.
Throughout the evolution quantum jumps occurring are accounted for by applying the jump operators $L$ directly to the states, i.e., $\ket{\phi'} = L \ket{\phi}$. 

In this method, we need an observable to measure how close to the steady state of the Liouvillian (or equivalently, the ground state of the parent Hamiltonian) the state is.
For this, we choose the excited population fraction $\nu = \langle \mathcal{N}_+ \rangle / \langle \mathcal{N} \rangle $, which approaches zero when the states reach the steady state.
After some initial transient decay, the average number of particles excited from the ground state (lower band of $H$) decreases exponentially to zero as $\nu(t) \propto\exp\left(-t / \tau\right)$, from which we may infer the cooling time $\tau$ by a numerical fit.
To estimate the statistical error of this fitting procedure, we employ the following bootstrap technique: From the $M=\num{10000}$ total samples, we infer the estimated value $\hat \tau$; to estimate the accuracy of $\hat \tau$, we draw $s=\num{40}$ smaller resamples of size $m = \num{1000} < M$ with replacement and compute the estimator $\tilde \tau_s$ for each.
Then, from the standard deviation of the $\tilde \tau$, we estimate the error of $\hat \tau$.
Additionally, by examining the individual trajectories, this method would also reveal potential instabilities in the model, where a macroscopic portion of them would fail to equilibrate.

\subsection{Truncated Hilbert Space}\label{sec:methods:ths}

An alternative approach to capture the relevant dynamics governing the system is the construction of a truncated Hilbert space $\mathcal{H}_{\mathrm t}$ in which the most relevant degrees of freedom are preserved~\cite{albert2025truncatedgaussianbasisapproach}.
The elementary excitations of the target Hamiltonian in \cref{eq:common_hamiltonian} are defects with one hole created by $x_-$ and a particle created by $x_+^\dagger$.
Let the Hilbert space with $0 \leq m \leq L$ such defects be $\mathcal H^{(m)}$, which is spanned by the orthonormal states 
\begin{align*}
    \ket{i_1 \ldots i_m; j_1\ldots j_m} = \prod_k x_{+,i_k}^\dagger x_{-,j_k} \ket{\circ}.
\end{align*}
The indices $i$ and $j$ denote the positions of excitations and holes, respectively.
Note that $\dim \mathcal H^{(m)} = \binom{L}{m}^2$, and thus the total Hilbert space may be partitioned as $\mathcal H = \bigotimes_{m} \mathcal H^{(m)}$~\footnote{This follows from a counting argument noticing that $\sum_{m=0}^L \binom{L}{m}^2= \binom{2L}{L} = \binom{N}{L} = \dim \mathcal H$.}.
As the late-time dynamics of the cooling protocol are dominated by states containing only few defects, we can limit our calculations to the truncated Hilbert space up to $M$ defects defined by
\begin{align*}
    \mathcal H_{\mathrm t;M} = \bigotimes_{m=0}^M \mathcal H^{(m)}.
\end{align*}
For $M \ll L$, this Hilbert space is now much smaller with $\dim \mathcal H_{\mathrm t;M} = \mathcal O(L^{2M})$.
In practice, $M=1,2$ already sufficiently captures the relevant time scales for the system sizes accessible.
As the truncated Hilbert space only captures a fraction of the full dynamics, it will generically provide a lower bound on the cooling time for a pure steady state given that there are no additional stationary solutions.

\section{Application to Cooling Times}\label{sec:results}

Using the numerical methods above, we discuss the cooling dynamics for the thermal dissipation in \cref{sec:results:thermal} and for the engineered protocol in \cref{sec:results:engineered}.

\subsection{Thermal Dissipation}\label{sec:results:thermal}

\begin{figure}
    \centering
    \includegraphics{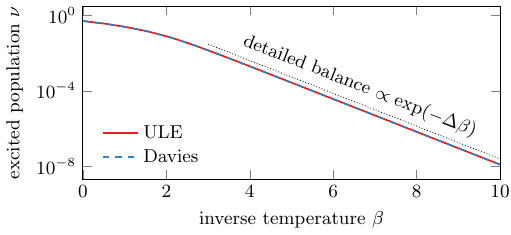}
    \caption{The excited population fraction $\nu = \tr (\rho_\infty \mathcal{N}_+) / \langle \mathcal{N} \rangle$ in the steady state $\rho_\infty$ for both thermal protocols at different inverse temperatures $\beta$ for system size $N=12$. Around $\beta \approx 4$, the excited population is effectively described by the detailed balance with respect to the level spacing $\Delta$ as enforced by the \gls{kms} condition in \cref{eq:kubomartinschwinger}. } \label{fig:cooling_targets}
\end{figure}

For the dynamics induced by weakly coupling the \gls{ssh} chain to the thermal bath, two distinct regimes emerge: the high-temperature regime with $\beta \approx 0$ and the low-temperature regime $\beta \gg 1$.
Only in the latter case is the system able to reach the ground state with substantial certainty, and thus most of the discussion will focus on $\beta \gg 1$, specifically $\beta = 10$.
In the following, we discuss these two distinct cases and touch on more generic setups obtained by changing the choice of system operators in \cref{eq:thermal_system_operators}.

\subsubsection{High-temperature regime}

As illustrated in \cref{fig:cooling_targets}, while the share of non-equilibrated particles $\nu$ in the steady state is the same for the Davies and \gls{ule} formulations, it is very heavily dependent on the choice of inverse temperature $\beta$.
While the drop begins subexponentially, at $\beta \gtrsim 4$ it becomes proportional to $e^{-\Delta \beta}$, where $\Delta$ is the difference between energy levels. 
The limiting behavior reflects the \gls{kms} condition on the spectral density function, cf.~\cref{eq:kubomartinschwinger}. 
Thus, while the thermal open system never \emph{fully} reaches the ground state at any finite temperature, it comes exponentially close to it. 
In our low-temperature discussion, we consider $\beta=10$ sufficiently close to zero temperature.
Note that the exact choice of $\beta$ does not qualitatively change the observations in the following, as the spectral density determining them is roughly the same for $\beta \gtrsim 5$, cf.~\cref{fig:spectral-density}.

\begin{figure}
    \centering
    \includegraphics{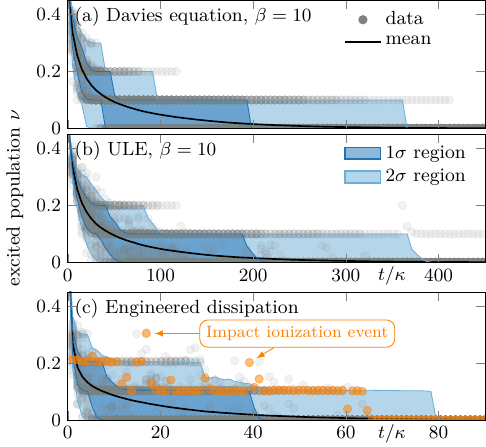}
    \caption{Comparison of the evolution of the excited population $\nu = \langle \mathcal{N}_+ \rangle / \langle \mathcal{N} \rangle$ for $m=\num{10000}$ quantum trajectories (of which $50$ are shown) for the three different protocols in the thermal (a) Davies and (b) \gls{ule} formulations as well as for (c) the engineered protocol. For each sample, the mean and $1\sigma$ ($2\sigma$) region containing $\SI{68.3}{\percent}$ ($\SI{95.4}{\percent}$) of the trajectories are shown. Clearly, the aggregate trajectory converges towards the targeted state with $\nu = 0$. Both the Davies and the \gls{ule} formulations yield virtually indistinguishable results. In all dissipation protocols, impact ionization events can be identified as illustrated for the orange trajectory in panel (c) which become increasingly insignificant at late times.}\label{fig:trajectories_comb}
\end{figure}

\subsubsection{Low-temperature regime}

\begin{figure}
    \centering
    \includegraphics{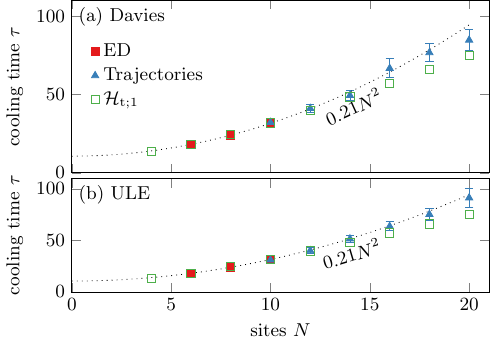}
    \caption{Dependence of the cooling time $\tau$ on the number of sites $N$ using site-switching system operators in \cref{eq:thermal_system_operators} at $\beta=10$. The data with error bars represent results from quantum trajectories. Both truncated Hilbert space ($\mathcal{H}_{\mathrm{t};1}$) and quantum trajectory calculations conform to a quadratic scaling law obtained by fitting to the exact data points.} \label{fig:sshulescaling}
\end{figure}

When setting $\beta = 10$ for the bath to allow the system to reach the ground state, by the \gls{kms} condition, almost only transitions decreasing the number of excitations are allowed, with the opposite direction being exponentially suppressed.
The processes governing the cooling in this protocol can be best understood from the realized quantum trajectories. 
Comparing them in \cref{fig:trajectories_comb}, we note two crucial insights:
First, the ensemble average between the Davies and \gls{ule} formulations is again approximately identical, confirming that the rotating wave approximation for the Davies  equation is well justified. 
Second, there are pronounced quantized plateaus at which the time evolution seems to be temporarily stuck.
These plateaus emerge, because the respective trajectory has few remaining excitations, which have yet to be dispersed.
That both the Davies and \gls{ule} formulation yield virtually indistinguishable dynamics can be confirmed by computing the cooling times as shown in \cref{fig:sshulescaling}.
There, both are found to be consistent with the same quadratic scaling in system size, which is also well captured by the truncated Hilbert space in first order~\cite{SuppMatFootnote}.

\subsubsection{On the concrete form of the thermal equation}

To verify that the diffusive behavior $\tau \propto N^2$ holds irrespective of our arbitrary choice of system operators, we implement a very asymmetric choice for the directionality of these cooling hoppings by removing all $X^\rightarrow$ and the $X_b^\updownarrows$ operators.
This then again produces a quadratic scaling of the cooling time $\tau \propto N^2$ while decreasing the proportionality constant to adjust for the impeded mobility of the excitations, cf.~\cref{fig:cooling_times_leftonly} in \cref{app:different_choices}. 
Similarly, other choices of spectral density functions, fulfilling the condition in \cref{eq:kubomartinschwinger}, also lead to a quadratic increase of the cooling time with system size. 
We replicate therefore the findings of \refcite{Chen2025} in demonstrating that, in general, thermal excitations only need polynomial time to cool a fairly general class of systems to their ground state.  

\subsection{Engineered Dissipation}\label{sec:results:engineered}

Qualitatively, the quantum trajectories observed in the engineered protocol are very similar to the low-temperature thermal coupling trajectories, cf.~\cref{fig:trajectories_comb}.
In particular, the engineered dissipation protocol features the same distinctive excited population plateaus.
When on a plateau, the excited population is decreased by annihilation of defects occurring as particle-hole excitations on top of the target state. However, until the individual trajectories are fully equilibrated, the excited population $\nu$ may occasionally increase, which is caused by \emph{impact ionization} events. They are proposed as a physical picture for an instability scenario: A hopping particle could reexcite an already equilibrated one, thus creating additional defects in the state~\cite{PhysRevB.107.174312,Lyublinskaya_2023,Lyublinskaya_2025,PhysRevB.107.174312,baburin2025minimalvelocitytravellingwave}. In that case, an instability could be rationalized in a scenario similar to absorbing state transitions~\cite{Kamenev_2011,Taeuber_2014}, where, in addition to a pure absorbing state, there is another mixed \enquote{active} state, which is the stable stationary state in the thermodynamic limit. While such impact ionization events certainly occur in both the thermal and the engineered protocol, their overall effect on the dark state population vanishes at late times, cf.~\cref{app:impact_ionization}.

\begin{figure}
    \centering
    \includegraphics{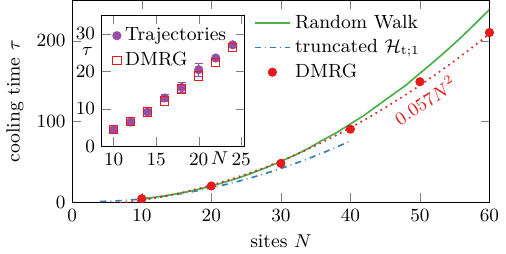}
    \caption{Dependence of the cooling time $\tau$ on system size $N$ for the engineered dissipation protocol in \cref{eq:engineered_jump_operators}. The inset confirms the agreement between cooling times obtained from \gls{dmrg} and quantum trajectory simulations with $m=\num{10000}$ samples ($m=\num{2000}$ samples at $N=22, 24$ with no error bars shown). For larger systems, the truncated Hilbert space provides a lower bound on the \gls{dmrg} data as expected, while the random-walk model slightly overestimates the required cooling time. All numerical simulations are consistent with a quadratic dependence $\tau \propto N^2$ indicating diffusive cooling. None of the methods show an instability of the steady state. \gls{dmrg} results have been verified using spectral gradient search up to $\SI{2}{\percent}$ deviation in the spectral gap.}
    \label{fig:sshengsites}
\end{figure}

\begin{figure}
    \centering
    \includegraphics{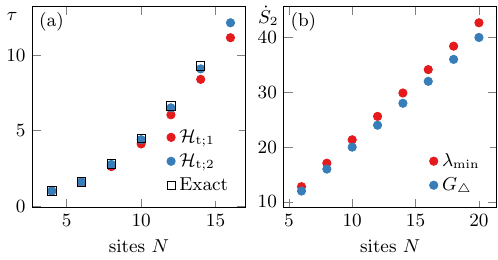}
    \caption{In panel (a), we show the convergence of the truncated Hilbert space approach towards the exact cooling time $\tau$. At $L=2$ the truncated Hilbert space is $\SI{100}{\percent}$ ($\SI{100}{\percent}$) and at $L=6$ is $\SI{4}{\percent}$ ($\SI{28}{\percent}$) of the total Hilbert space in first (second) order. It is evident that the qualitative behavior is already captured reasonably well at first order. In panel (b) the entanglement growth rate $\dot S_2$ predicted for the fully excited state $\ket{\triangle}$ as measured by $G_\triangle$ is compared to the maximum possible given by the minimal eigenvalue $\lambda_{\min}$ of $\mathcal H = \mathcal L + \mathcal L^\dagger$. Both grow linearly in system size and agree almost perfectly.} \label{fig:engineered_trunchilbertspace_entanglementgrowth}
\end{figure}

In \cref{fig:sshengsites}, we plot the cooling times obtained from quantum trajectories, \gls{ed}, and \gls{dmrg} results, finding perfect agreement in their overlapping regimes.
We also find them compatible with the truncated Hilbert space results~\cite{SuppMatFootnote}.
As predicted by the random-walk model derived in \cref{sec:diffusion:random-walk}, the cooling time scales quadratically with system size, although the exact diffusion constant disagrees slightly.
Specifically for the truncated Hilbert space, in \cref{fig:engineered_trunchilbertspace_entanglementgrowth}\,(a) we compare the convergence of the truncated Hilbert space approach in both first and second order to the exact results.
While inclusion of the second order improves the quantitative accuracy, the qualitative behavior is already captured remarkably well by considering only one excitation-hole pair.

In the upcoming section, we will present a purely classical description of the dynamics, where a random-walk of the defects leads to diffusive cooling. In this particle picture, we expect ballistic entanglement growth on short time scales~\cite{Calabrese_2005}.
This growth is the fastest if all sites are excited, because then the number of entanglement generating quasiparticles is the highest. 
We can measure the entanglement by the Rényi $\alpha=2$ entropy defined as 
\begin{align*}
    S_2(\rho)=-\ln \tr \rho^2 = -\ln \langle  \rho, \rho \rangle,
\end{align*}
where $\langle X, Y\rangle=\tr X^\dagger Y$ is the Hilbert Schmidt inner product.
If the state $\rho$ is governed by the Liouvillian, we find the time evolution of the Rényi entropy as $\dot S_2 = -\langle \rho, \mathcal H[\rho]\rangle / \langle \rho, \rho \rangle$ with the symmetrized Liouvillian $\mathcal H = \mathcal L + \mathcal L^\dagger$.
This object is Hermitian and thus accessible by \gls{dmrg} methods.
Indeed, we can bound the entanglement growth rate by the minimal eigenvalues, i.e., $\dot S_2 \leq -\lambda_{\min}$, and thus gain insights into the short-time dynamics occurring in the system away from the steady state. 
We expect linear entanglement growth $\lambda_{\min} \propto L$ in short time scales for the ballistic mode~\cite{_nidari__2020,Calabrese_2005}.
Let $\ket{\triangle} = \prod_j x_{j,+}^\dagger \ket{0}$ be the state where all bonds are excited; we define the corresponding density matrix $\rho_\triangle=\ketbra{\triangle}$ and the entanglement growth $G_\triangle=-\langle \rho_\triangle, \mathcal H[\rho_\triangle]\rangle$ in the state.
In \cref{fig:engineered_trunchilbertspace_entanglementgrowth}\,(b), we show both $G_\triangle$ and the minimal eigenvalue $\lambda_{\min}$ of $\mathcal H = \mathcal L + \mathcal L^\dagger$, confirming that both are very close and linearly dependent on the system size.

\section{Diffusion and Random-Walks}\label{sec:diffusion}

\begin{figure*}
    \includegraphics{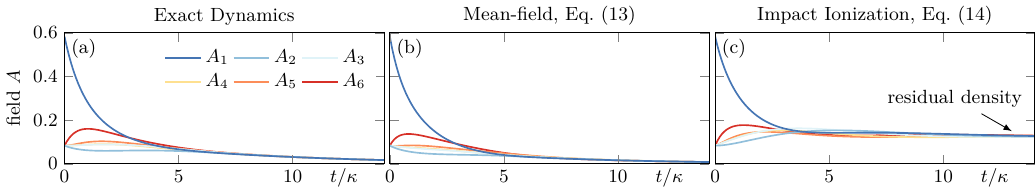}
    \caption{In panel (a), we show the time evolution of the fields $A$ on different sites starting from the state $\ket{\iota_6}$ for $N=12$. This is compared in panels (b) and (c) with the solution of the field equations in \cref{eq:ABfields,eq:impact_ionization} (with $\tau_\pm=1$), respectively. While \cref{eq:ABfields} captures the short-time dynamics qualitatively very accurately and by the finite-size correction shows the correct decay in the long-time limit, the impact ionization present in \cref{eq:impact_ionization} would predict a residual density deviation, which is not observed numerically.} \label{fig:Afieldsmaintext}
\end{figure*}

The previous section extensively benchmarked the cooling rates in the engineered and thermal protocols, finding both to be quadratic in the regimes reaching the ground state, with $\tau \propto N^2$, indicating diffusive behavior associated with total particle number conservation.
However, the numerical data does not by itself provide a clear explanation of the quadratic cooling time observed.
In this section, we focus on the engineered dissipation and introduce a simple model explanation of its quadratic scaling by (i) deriving a mean-field approximation and (ii) mapping the cooling to random-walks of the excitations on the wires.

\subsection{Mean-field Description}\label{sec:diffusion:mean-field}

Already on a mean-field level, the diffusive nature of the excitations and thus the quadratic scaling of the cooling time with system size can be derived, with the following condensed derivation being complemented by more details provided in \cref{app:engineered_mean_field}.
Starting from the Heisenberg equation of motion for an operator $O$ giving its time-evolution 
\begin{align*}
    \partial_t O = \mathcal{L}^\dagger [O] = -i[O, H] + \sum_{\mu} \Big(L_\mu^\dagger O L_\mu - \frac{1}{2} \{L_\mu^\dagger L_\mu, O\}\Big),
\end{align*}
we derive the equations of motion for the density correlators $n_{j,\sigma\sigma'} = x_{j,\sigma}^\dagger x_{j,\sigma'}$, cf.~\cref{eq:eom_full_npp,eq:eom_full_nmm,eq:eom_full_npm,eq:eom_full_nmp}.
These equations of motion are then turned into ones for the expectation values $\langle n_{j,\sigma\sigma'} \rangle$ for late times at which the system is approximately Gaussian.
Introducing $a$ as the lattice spacing of the model and taking the continuum limit $a\to 0$ then allows us to move to continuum excitation densities $n_{\sigma\sigma'}(x=ja) = \langle n_{j,\sigma\sigma'} \rangle$.
A compact statement of our results is---in light of previous results for similar dissipative properties derived from Keldysh field theory~\cite{Lyublinskaya_2025}---given in terms of the following quantities:
Concentrating on the deviations of the two excitation densities from the dark state, we define $\delta n_+ = \langle n_{++} \rangle$ and $\delta n_- =  1 - \langle n_{--} \rangle$, such that $\alpha = (\delta n_+ + \delta n_-)/2$ and $\beta = (\delta n_+ - \delta n_-) / 2$ are the total density deviation and the density imbalance, respectively.
To cancel a drift term in the following equation and allow for better comparison to the literature, we define $Q(x) = q(x - vt)$ with $Q=A,B$, $q=\alpha,\beta$, and $v= -\kappa a/2$. 
For the engineered dissipation, we find that $A$ and $B$ abide by the reaction diffusion (in the late-time limit with $A, B \ll 1$; cf.~\cref{eq:AB:1,eq:AB:2} for the full model)  
\begin{align}
    \left(\partial_t - D_+ \partial_x^2\right) A + \frac{\beta}{2} A^2 + \abs{\lambda_L} A &= D_- \partial_x^2  B + \frac{\beta}{2} B^2, \nonumber\\
    D_- \partial_x^2 A &= \left( \partial_t  - D_+ \partial_x^2  \right) B, \label{eq:ABfields}
\end{align}
with coefficients 
\begin{align*}
    D_+ &= \frac{\kappa a^2}{4}, & D_- &= 0, & \beta &= 4\kappa,
\end{align*}
and $\lambda_L$ the Liouvillian gap introduced for numerical comparisons to finite systems vanishing in the thermodynamic limit, i.e., $\lambda_L \rightarrow 0$ for $L\rightarrow \infty$.

Keldysh field theory calculations on similar models yielded the coupled \gls{fkpp} equations~\cite[Eq.~(26)]{Lyublinskaya_2025} 
\begin{align}
    (\partial_t - D_+ \partial_x^2 - \frac{1}{\tau_+}) A + \frac{\beta}{2} A^2 &= (D_- \partial_x^2 + \frac{1}{\tau_-}) B + \frac{\beta}{2} B^2, \nonumber\\
    D_- \partial_x^2 A &= (\partial_t - D_+ \partial_x^2) B. \label{eq:impact_ionization}
\end{align}
Crucially, these differ from our findings by additional impact ionization terms $\tau_\pm$, which, if present, render the dark state unstable by allowing for nonvanishing total density deviations in the long time limit.
These terms can, however, be ruled out for the engineered protocol presented here.
To substantiate this, we prepare the system in the state 
\begin{align}
    \ket{\iota_n} &= \frac{1}{\sqrt{n}} \Big(\sum_{j=1}^n x_{j,+}^\dagger\Big) x_{1,-} \ket{\circ} \label{eq:iotastate}
\end{align}
for which we expect a large contribution of the impact ionization, if any is present.
In \cref{fig:Afieldsmaintext}, we show the time evolution of the fields $A$ at different sites compared to the diffusion equation in \cref{eq:ABfields} as well as those with the same coefficients in \cref{eq:impact_ionization} with a finite ionization constant $\tau_+=\tau_- = 1$.
Clearly, there is no residual density deviation from the steady state as would be a consequence of the destabilizing impact ionization.

\subsection{Random-Walk Model}\label{sec:diffusion:random-walk}

In the previous sections, we investigated the full quantum description of the cooling protocols and provided mean-field equations capturing the dynamics of relevant observables. While these two approaches hint at diffusive dynamics governing the cooling, they lack an intuitive justification. In the following, we derive a classical random-walk model underpinning the diffusive character of the cooling protocols.
Starting from the engineered jump operators in \cref{eq:engineered_jump_operators}, we rewrite the local wire operators as 
\begin{align*}
    c_{j+1,a} &= \frac{1}{\sqrt{2}} (x_{j,+} - x_{j,-}), &
    c_{j,b} &= \frac{1}{\sqrt{2}} (x_{j,+} + x_{j,-}), 
\end{align*}
such that the jump operators read (omitting the proportionality constant $\sqrt{\kappa/2}$ to improve readability) 
\begin{align*}
    L^+_{j,a} &\propto (x_{j-1,+}^\dagger - x_{j-1,-}^\dagger) x_{j,+}, &
    L^+_{j,b} &\propto (x_{j,+}^\dagger + x_{j,-}^\dagger) x_{j,+}, \\
    L^-_{j,a} &\propto (x_{j-1,+} - x_{j-1,-}) x^\dagger_{j,-}, &
    L^-_{j,b} &\propto  (x_{j,+} + x_{j,-}) x^\dagger_{j,-}.
\end{align*}
Remarkably, these jump operators are local for both particles and holes and only decrease the number of defects in a state, as there is no term of the form $x_+^\dagger x_-$.
Thus, in the late stage of the evolution, there will eventually be only one particle $x_{j+}^\dagger$ and one hole $x_{i-}$ in the system, which hop around as mediated by the jump operators. 
While the operators $L_{b}^\pm$ act only on site, it is  $L_a^+$ ($L_a^-$) that allows the excitation (hole) to jump left.
Once both are on the same site, they can annihilate, thus equilibrating the system to the steady state.

\begin{figure}
    \includegraphics{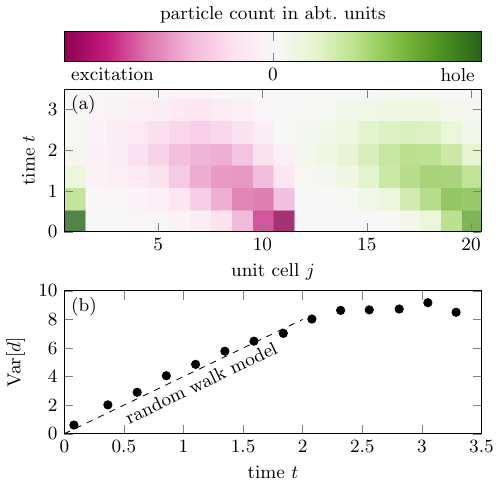}
    \caption{Illustration of the random-walk dynamics exhibited by the defects, where in panel (a) we initialize hole (excitation) in unit cell $i_0=1$ ($j_0=11$) in a system with $N=40$ sites and show the histogram of their positions for $1000$ instances of a quantum trajectory in the engineered dissipation scenario. Both travel to the left and show a diffusive cone. In panel (b), the variance of the position difference $d=i - j$ between their positions, showing perfect linear growth in the beginning, as expected from the random-walk model.}\label{fig:random_walk}
\end{figure}

In \cref{fig:random_walk} we show the time evolution of a state initialized in such a late-time configuration with the position of the hole and excitation averaged over multiple trajectories.
There, it can be clearly seen that both move to the left only.
We also compare to the analytical solution of the model derived in the following.

\subsubsection{Analytical Prediction of Cooling Rates}

The operators $A_\mu = L_\mu^\dagger L_\mu$ (with $\mu$ enumerating all Lindblad operators) forming the terms of the effective Hamiltonian $\tilde H = -\frac{i}{2} \sum_\mu A_\mu$ are 
\begin{align}
    A^+_{j,\gamma} \equiv (L_{j,\gamma}^+)^\dagger L_{j,\gamma}^+ &= \kappa x_{j,+}^\dagger (1-n_{j,\gamma}) x_{j,+}, \nonumber\\
    A^-_{j,\gamma} \equiv (L_{j,\gamma}^-)^\dagger L_{j,\gamma}^- &= \kappa x_{j,-} n_{j,\gamma} x_{j,-}^\dagger, \label{eq:Amuoperators}
\end{align}
where $\gamma=a,b$ the wire index.
Thus, the exponential damping $\exp(-i\tilde Ht)$ does not change the state but merely decreases its amplitude.
Consider for now a trajectory in which only one defect remains, given by the state $\ket{i;j} \in \mathcal{H}^{(1)}$ with $i$ and $j$ sufficiently far apart such that no operators in the following have overlapping support.
There are four nonvanishing  operators in the effective Hamiltonian, namely $A^+_{i,\gamma}$ and $A^-_{j,\gamma}$, $\gamma=a,b$, whose action on the state is 
\begin{align*}
    A^\pm_{i/j,\gamma} \ket{i;j} = \frac{\kappa}{2} \ket{i;j}.
\end{align*}
The norm of the time evolved state $\ket{\psi(0)} = \ket{i;j}$ is therefore given by 
\begin{align*}
    \norm{\psi(t)} = \exp\left( -\kappa t \right).
\end{align*}
A quantum jump occurs for $\norm{\psi(T)}^2 = r$. with $r$ a uniform random variable; thus, the jump time $T$ is exponentially distributed with mean $1/2$, i.e., two jumps occur per unit time. 
Notably, the jump frequency is also independent of system size.
Once the jump occurs, the jump operators act on the state as follows: 
\begin{align*}
    L_{i,a}^+ \ket{i;j} &\propto \ket{i-1;j}, & 
    L_{j,a}^- \ket{i;j} &\propto \ket{i;j-1}, \\ 
    L_{i,b}^+ \ket{i;j} &\propto \ket{i;j}, & 
    L_{j,b}^- \ket{i;j} &\propto \ket{i;j},
\end{align*}
with the same proportionality constant in each term.
Thus, they are equally likely to occur and the random-walk in the difference $d = i - j$ is changed as $d \rightarrow d\pm 1$ with probability $1/4$ and remains unchanged with probability $1/2$.
The hitting time of this random-walk, i.e., the number of steps $n$ at which it reaches the boundary $L/2$ is 
\begin{align*}
    E[n] = 2 (L/2)^2 = L^2/2.
\end{align*}
By plugging in the relationship between number of steps and time $t$, we find the cooling time 
\begin{align}
    \tau = \frac{n}{2} = \frac{L^2}{4}. \label{eq:random_walk}
\end{align}
Note that here we assumed that the hole-excitation pair starts of at the worst possible position.
Numerically, we found the actual proportionality to be $\tau \approx 0.225 L^2$.

\subsubsection{Generalized Model}

\begin{figure}
    \includegraphics{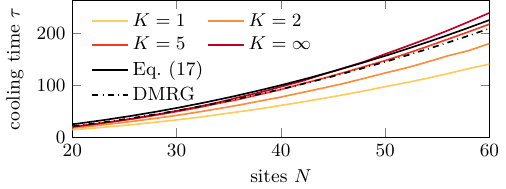}
    \caption{The cooling time $\tau$ predicted by the more sophisticated random-walk model run for $\num{100000}$ samples with the maximum number of excitations being $K$, i.e., $K=\infty$ means there are no restrictions. With more excitations included, the cooling time increases. We include the fit obtained from the \gls{dmrg} data (dash dotted) from \cref{fig:sshengsites} as well as the prediction in \cref{eq:random_walk}. Overall, the random-walk model predicts longer cooling times than the quantum simulation, which can be attributed to the fact that it is both classical and strictly local.} \label{fig:sophisticated_random_ralk}
\end{figure}

While the above random-walk  model correctly captures the classical dynamics giving rise to the diffusive cooling, it is restricted to only one defect living in $\mathcal{H}^{(1)}$. Furthermore, it makes ad hoc assumptions about the initial spatial extent of the defect. In the following, we provide a generalized model that relaxes these assumptions while finding close quantitative agreement with the fully quantum description. For this, we allow for a state with multiple defects and implement the following three improvements: (i) inclusion of the decay rates given by $A^\pm$ when excitation and hole overlap, (ii) a correct treatment of the jump operators' actions when hole and excitation are close together, and (iii) a way to initialize an appropriately distributed random initial state with multiple particle-hole pairs.
The details outlined below yield the recipe by which the results in \cref{fig:sophisticated_random_ralk} are obtained.
While the quadratic dependence $\tau \propto N^2$ matches the findings for the quantum formulation, the generalized random-walk model quantitatively predicts a marginally higher diffusion constant than measured for the numerical data in \cref{fig:sshengsites}.
This can mostly be attributed to the neglect of off-diagonal terms as outlined below, and the fact that while $\exp(\mathcal{L} t)$ can generate nonlocal terms in its Taylor expansion, the random-walk is limited to local jumps only.

To derive the model, note that the wire number operators $n_{j,a/b}$ can be written in the $x$ basis in terms of their number operators $n_{j,\pm} = x_{j,\pm}^\dagger x_{j,\pm}$ as 
\begin{align*}
    n_{j+1,a} &= \frac{n_{j,+} + n_{j,-} - Y_{j}}{2}, & n_{j,b} &= \frac{n_{j,+} + n_{j,-} + Y_j}{2},
\end{align*}
having defined $Y_j \equiv x_{j,+}^\dagger x_{j,-} + x_{j,-}^\dagger x_{j,+}$.
We obtain the following expressions for the $A^\pm$ operators:
\begin{align*}
    A^+_{j,a} &= n_{j,+} (1 - n_{j,a}), &
    A^+_{j,b} &= n_{j,+} (1 - n_{j,-} / 2), \\
    A^-_{j,a} &= (1 - n_{j,-}) n_{j,a}, &
    A^-_{j,b} &= (1 - n_{j,-}) (1 + n_{j,+}) / 2.
\end{align*}
Unfortunately, the operators with $\gamma=a$ are not diagonal in the $x^\pm$ basis, such that we need to employ the mean-field approximation $\langle Y_j \rangle \approxeq 0$, with which
\begin{align*}
    n_{j,a} \approxeq \frac{n_{j-1,+} + n_{j-1,-}}{2}.
\end{align*}
Having computed the $A_\mu$, we randomly choose the jumps in the same way as in the quantum jump model prescription.  
To find the initial distributions, we weigh the initial number of defects $m$ by the respective dimensions of their Hilbert spaces $p_m \propto \dim \mathcal{H}^{(m)}$ up to normalization.
The particle-hole pairs themselves are then placed randomly on the wires without repetition.

\subsection{Thermal Dissipation}\label{sec:diffusion:thermaldissipation}
It might seem counterintuitive that this same reasoning should apply for the thermal dissipation in the low-temperature regime as well. 
However, the structure of the Lindblad jump operators is very similar in nature to the engineered protocol if the temperature $T \ll \Lambda_0$ ($\beta\Lambda_0 = 10 \gg 1$) is low. 
Both share particle number conservation and when the system operators in \cref{eq:thermal_system_operators} are written in the basis of the bond operators $x_\pm$, they contain terms of the following form~\footnote{The full description of the system operators in this basis is provided in the Supplementary Material.}:
\begin{align*}
    J_{ij}^{(1)} &= \frac{\gamma_0}{2} x_{i,+}^\dagger x_{j,+}, &
    J_{ij}^{(2)} &= \frac{\gamma_0}{2} x_{i,-}^\dagger x_{j,-}, \\
    J_{ij}^{(-)} &= \frac{\gamma_-}{2} x_{i,+}^\dagger x_{j,-}, &
    J_{ij}^{(+)} &= \frac{\gamma_+}{2} x_{i,-}^\dagger x_{j,+}, 
\end{align*}
where we defined $\gamma_0 = \sqrt{J(0)}$, $\gamma_\pm = \sqrt{J(\pm2t)}$.
Here, $i$ and $j$ are neighboring sites, rendering all the hopping operators $J$ quasi local.
In the low-temperature limit, the operator $J^{(-)}$ gets exponentially suppressed with $\gamma_- \ll 1$, while the other combinations retain approximately similar prefactors $\gamma_0 \approx \gamma_+ = \mathcal O(1)$.
Thus, at low-temperatures, the Lindblad operators also never increase the number of defects.
Furthermore, no jumps are mediated by the Hamiltonian in \cref{eq:common_hamiltonian}, which only introduced local phases for states expressed in the particle-hole picture.
One can therefore identify a similar cooling mechanism as with engineered dissipation, where holes and excitations are moved locally by the jumps until they are next to one another, upon which a jump triggers that recombines excitation and hole. 

\section{Comparison to analytical work}\label{sec:comparison}

A series of recent analytical works analyzed the late-time dynamics of dissipatively induced topological insulators~\cite{PhysRevB.107.174312,Lyublinskaya_2023,Lyublinskaya_2025,baburin2025minimalvelocitytravellingwave} and superconductors~\cite{shustin2025dissipation}. 
These works build on a Keldysh field theory approach for continuum models in one and two dimensions to compute the effective action for the particle and hole fluctuations on top of a dissipative topological insulator beyond the mean field approximation~\cite{PhysRevLett.124.240404}, subject to the overall particle number constraint. 
This approach describes the near-stationary properties of the system. It finds a diffusive regime dominated by the conservation of total particle number, followed by an asymptotic behavior governed by coupled \gls{fkpp} equations, where residual particles and holes can annihilate~\cite{PhysRevB.107.174312,Lyublinskaya_2023,Lyublinskaya_2025}. 
In particular, these works predict an instability of the dark state towards a distinct, mixed stationary state, based on the perturbative calculation of the coefficients of the \gls{fkpp} equations. 
In Ref.~\cite{baburin2025minimalvelocitytravellingwave}, it is argued that two coupled \gls{fkpp} equations generically show stable and unstable solutions in some analogy to impact ionization.

Our present numerical study in one dimension is not compatible with these findings. 
We find stable dark state solutions in all parameter regimes considered, including in parameter regimes that reduce---in the continuum limit---to models analyzed in the above works. 
Moreover, in \cref{app:uniqueness} we offer an analytical proof of the uniqueness of the dark state for models in the class considered here and in previous literature~\cite{Huang_2022,mittal2025fermionquantumcriticalityfar,PhysRevLett.124.240404,PhysRevResearch.3.043119,Altland2021}; uniqueness excludes an instability towards an alternative stationary state. 
Our results are also in line with the stability of dissipatively induced topological superconductors established numerically in Ref.~\cite{Iemini_2016}.
In particular, this shows not only that no other dark states exist, but also that no other mixed stationary solutions exist for any system size, including in the thermodynamic limit $N \to \infty$. 
Within a mean-field description that allows for the description of nonlinear deterministic dynamics, we find that the dynamics are best described by coupled diffusion instead of \gls{fkpp} equations.
Altogether, this speaks against the impact ionization scenario also on analytical grounds. 
In addition, we test the field equations numerically, confirming our analytical expectations.

\section{Concluding Discussion}\label{sec:discussion}

Our results demonstrate that the number-conserving ground-state preparation of a dimerized \gls{ssh} Hamiltonian using either local thermal cooling or a dissipative state preparation protocol is governed by diffusive processes that equilibrate the system in time quadratically dependent on system size.
In neither protocol we found obstructions fundamentally hindering their cooling capabilities. In the case of the engineered protocol, additional stationary solutions of the dynamics next to the target state pose such obstacles to cooling. However, for the dissipative state preparation protocol studied in this work, we provided proof of the uniqueness of the steady state. For thermal cooling, local minima may hinder equilibration. Relying on finite size-scaling of the cooling time, we found no indication of the cooling process getting trapped away from the target state in either protocol.
The latter is by no means guaranteed, as Nature is constrained to efficiently finding local minima~\cite{Chen2025}, while the ground state is---by definition---a global minimum.
When other local minima are present, they may hamper thermal preparation of the ground state, in which case we would expect exponential scaling of the cooling times in system size.
From a complexity theoretical perspective, the quadratic scaling of cooling time found in this work reflects the fact that the parent Hamiltonian is quantumly easy~\cite{Chen2025} and that the considered system-bath couplings are sufficiently ergodic.

Comparing the two protocols, we found that the thermal protocol takes longer to cool than the engineered one.
While this finding quantitatively depends on the choice of parameters and a priori assumptions on the dynamics, we expect more favorable prefactors (diffusion constants) of the quadratic relaxation time-scale to also manifest in experimentally relevant platforms.
Microscopically, descriptions of thermal coupling in terms of master equations for our model are derived within the Born-Markov approximation~\cite[Chap.~3.3]{Breuer2007}. 
This weak coupling assumption constrains the cooling rates to be smaller than the system energy scales, which may be as low as $E/\hbar = \omega \sim \si{kHz}$ in atomic quantum simulators~\cite{PRXQuantum.2.017003}.
Equilibration then necessitates extremely low bath temperatures $T < \hbar\omega/k_B \approx \SI{10}{\nano\kelvin}$ and, relating to our present findings, is delayed by small diffusion constants. This is contrasted by the engineered dissipation setting, where the dissipative processes typically involve excitations to energetically higher-lying internal states, e.g., optical excitations of atoms faster than terahertz that may then dissipatively relax by spontaneous emission~\cite{Diehl2008}. The Born-Markov approximation is then assumed with respect to these faster timescales  thus allowing for significantly higher coupling rates in the resulting Lindblad master equation.

\section*{Data Availability}

The data that support the findings of this article are openly available on Zenodo~\cite{pokart_2026_18379558}.

\acknowledgments

The authors are grateful to N.~Albert, I.~Burmistrov, A.~Kamenev, J.~Lang, R.~Mittal, J.~Schwardt, and F.~Thompson for insightful discussions.
Calculations were performed using \package{ITensors.jl}~\cite{10.21468/SciPostPhysCodeb.4,10.21468/SciPostPhysCodeb.4-r0.3}, \package{DifferentialEquations.jl}~\cite{DifferentialEquations.jl-2017}, \package{KrylovKit.jl}~\cite{haegeman_2025_17316953}, \package{Roots.jl}~\cite{Roots.jl} and \package{Numpy}~\cite{harris2020array}.
{T.P.}, {L.K.} and {J.C.B.} acknowledge financial support from the Deutsche Forschungsgemeinschaft (DFG, German Research Foundation) through the Collaborative Research Centre SFB 1143, the Cluster of Excellence ct.qmat, and the DFG Project No.~419241108. 
{S.D.} is supported by the DFG under Germany's Excellence Strategy Cluster of Excellence Matter and Light for Quantum Computing (ML4Q) EXC 2004/1 390534769 and by the DFG Collaborative Research Center (CRC) 183 Project No.~277101999 -- project B02.
The authors gratefully acknowledge the computing time made available to them on the high-performance computer at the NHR Center of TU Dresden. This center is jointly supported by the Federal Ministry of Education and Research and the state governments participating in the NHR.

\bibliography{sources}

\appendix
\onecolumngrid

\section{Example of an unstable steady state}\label{app:uniqueness:second_dark_state}

In previous works discussing the instability scenario, cartoon examples of unstable dissipative state preparation protocols were constructed~\cite{Lyublinskaya_2023,Lyublinskaya_2025,PhysRevB.107.174312}.
Here, we compare these previous works to our model and explicitly construct the additional dark state of the protocol in Ref.~\cite[App.~D]{PhysRevB.107.174312}, which in our notation is given by the Lindblad operator
\begin{align*}
    L^1_j = L_{j,a}^+ &= \sqrt{\frac{\kappa}{2}} (x_{j-1,+}^\dagger - x_{j-1,-}^\dagger) x_{j,+}.
\end{align*}
While this model still has $\ket{\circ}$ as its dark state, we may also construct the following vector.
For this, define 
\begin{align*}
    \ket{\phi_n} = \sum_{\vec k} (x_{k_1,+}^\dagger x_{k_1,-}) \cdots (x_{k_n,+}^\dagger x_{k_n,-}) \ket{\circ},
\end{align*}
where the sum is over all vectors $\vec{k}=(k_1, \ldots, k_n)$ with $n$ unique integer indices on the lattice.
The second dark state is then $\ket{\square} = \sum_n (-1)^n \ket{\phi_n}$. 
To see this, we apply $x_{m,-}^\dagger$ to the $\ket{\phi_n}$, finding 
\begin{align*}
    x_{m,-}^\dagger \ket{\phi_n} &= \sum_{\vec k} x_{m,-}^\dagger (x_{k_1,+}^\dagger x_{k_1,-}) \cdots (x_{k_n,+}^\dagger x_{k_n,-}) \ket{\circ} \\
    &= \sum_{\vec k} \begin{cases}
       -(x_{k_1,+}^\dagger x_{k_1,-}) \cdots (x_{k_j,+}^\dagger) \cdots (x_{k_n,+}^\dagger x_{k_n,-})\ket{\circ}, &\text{if}\ m = k_j, \\
       0, &\text{if}\ m \neq k_j,
    \end{cases}
\end{align*}
which is given more compactly by $x_{m,-}^\dagger \ket{\phi_n} = -x_{m,+}^\dagger \ket{\phi_{n-1}}$.
This yields the recursion relation 
\begin{align*}
    L^1_m \ket{\phi_n} = \sqrt{\frac{\kappa}{2}} (x_{m-1,+}^\dagger - x_{m-1,-}^\dagger) x_{m,+} \ket{\phi_n} = \sqrt{\frac{\kappa}{2}} x_{m-1,+}^\dagger x_{m,+} (\ket{\phi_{n}} + \ket{\phi_{n-1}}).
\end{align*}
Thus, summing the $\ket{\phi_n}$ with alternating signs cancels the effect of $L^1_m$.
This finding has been confirmed numerically, i.e., $\ket{\square}$ really is the second dark state.
For $L=2$, the state is 
\begin{align*}
    \ket{\square}=(x_{1+}^\dagger x_{2+}^\dagger - x_{1-}^\dagger x_{1+}^\dagger - x_{2-}^\dagger x_{2+}^\dagger)\ket{0}.
\end{align*}
As there are now two dark states, the steady state manifold is four-times degenerate, which can also be tested numerically.

\begin{figure}
    \centering
    \includegraphics[width=0.5\linewidth]{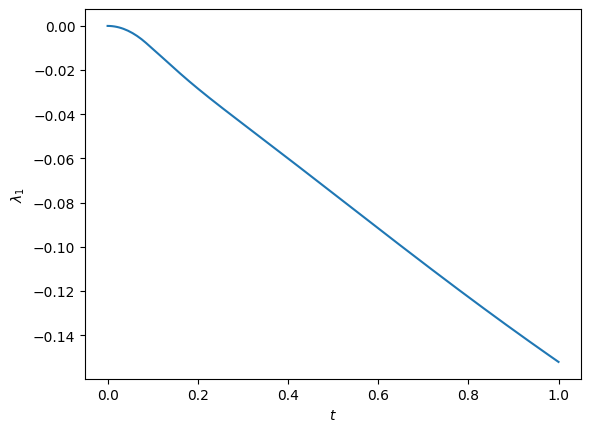}
    \caption{The first eigenvalue of the Lindbladian $L_{j,a}^+$ and $tL_{j,b}^+$. Here, $N=8$.}
    \label{fig:burmistrov_gap}
\end{figure}

Finally, we may check how stable this gap closing is.
For this, we consider the Lindblad operators $L_{j,a}^+$ and $tL_{j,b}^+$, where $0\leq t \leq 1$ allows us to tune the strength of the Lindblad operator destabilizing $\ket{\square}$.
See \cref{fig:burmistrov_gap} for the plot of the Liouvillian gap.
The gap closing is not stable with respect to the introduction of the additional terms, and therefore, only if all Lindblad operators except for $L^1$ are omitted, a non-unique and unstable situation leads to an instability scenario.

\section{On the Uniqueness of the Steady State}\label{app:uniqueness}

Proving that the dark state $\ket{\circ}$ is a unique dark state of the engineered dissipation given by the Lindblad operators in \cref{eq:engineered_jump_operators} is done using two separate ingredients~\cite{PhysRevA.78.042307}: (i) showing that the dark subspace is indeed one-dimensional, and (ii) proving that no other solutions outside the dark subspace exist.

The (i) one-dimensionality of the dark subspace can be seen immediately by construction. 
That is, the Hilbert space with no defects $\mathcal{H}^{(0)}$ is itself one-dimensional.
Next, we show that (ii) no other dark states exist, by proving that every state in the Hilbert space is connected to the dark state by a sequence of jumps, thus proving its uniqueness~\cite[Theorem 2]{PhysRevA.78.042307}.
The intuition is that such a sequence will appear in the Taylor expansion of $e^{\mathcal{L}t}$, and thus slowly drain the state into the dark state.
Write an arbitrary state $\ket{\psi}$ in the basis used to span the truncated Hilbert space in \cref{sec:methods:ths}.
That is, let 
\begin{align*}
    \ket{\psi} = \sum_{m=0}^L \sum_{\abs{\vec{i}}=\abs{\vec{j}}=m} \Psi^{(m)}_{i_1\ldots i_m;j_1\ldots j_m} \ket{i_1\ldots i_m;j_1\ldots j_m}.
\end{align*}
Here, $\vec{i}=(i_1,\ldots,i_k)$ and $\vec{j}=(j_1,\ldots,j_k)$ denote multi-indices indexing the unit cells of the system, where $\vec{i}$ ($\vec{j}$) are the positions of the excitations (holes), cf.~\cref{sec:methods:ths}. Their length is given by their number of entries $\abs{\vec{i}} = k$.
Now, suppose that $\Psi^{(0)}=0$, i.e., there is no overlap with the dark state.
We will now show that there exists a sequence of jumps that creates some finite overlap with the dark state.
To this end, let us concentrate on the operators 
\begin{align*}
    L^1_j\equiv L^+_{j,a} &\propto (x_{j-1,+}^\dagger - x_{j-1,-}^\dagger) x_{j,+}, &
    L^2_j\equiv L^-_{j,b} &\propto  (x_{j,+} + x_{j,-}) x^\dagger_{j,-}.
\end{align*} 
In the following, $L^1$ transports an excitation to the left by one cell while keeping the number of defects fixed (up until the last jump) and $L^2$ annihilates a defect, thus connecting the sector $m \rightarrow m-1$.
The idea will be to use the operator $L^1$ to bring excited particles spatially on top of the hole, i.e., to the same site $j$, and then use $L^2$ to annihilate the particle-hole pair. 
More precisely, now pick in $\ket{\psi}$ the sector with smallest $m$ such that there exists a $\Psi^{(m)} \neq 0$.
Within this sector, pick the states with the minimal distance $i_k - j_{k'}$ between a neighboring particle-hole pair; \emph{minimal} is defined with respect to the ability of the excitation to the move to the left only.
By construction of the minimal distance, there is no defect on the way that could block the following transport by the Pauli exclusion principle.
Then, apply the following move operator:
\begin{align*}
    M = L^2_{j_{k'}} \prod_{i=j_{k'}+1}^{i_k} L^1_{i}.
\end{align*}
The resulting wave function now has a nonvanishing component in the $m-1$ sector.
This then by induction implies that any state can be connected to the dark state, as the argument can be repeated until the dark subspace is reached. Because the dark state does---by definition---not couple to any of the jump operators, the argument also holds independently of the assumption $\Psi^{(0)} \neq 0$ and thus for arbitrary states. Intuitively, by this mechanism, the operators constitute a pump that drains all amplitudes into the dark subspace.
Importantly, note the difference if we omitted $L^2$: While this operator can still equilibrate an excitation, it has to by considering neighboring sites and thus is susceptible to Pauli blocking leading to the anomalous dark state constructed in \cref{app:uniqueness:second_dark_state}.

\section{Preconditioning and Stopping Spectral Gradient Search}\label{app:sgs_preconditioning}

The spectral gradient search in \cref{sec:methods:vectorization} benefits greatly from effective preconditioning, which speeds up the convergence toward the ground state of $X_{\tilde \lambda}$ in \cref{eq:sgs_xlambda}.
To this end, let $\ket{0}$ be the steady state and $\ket{1}$ be first excited state obtained from the previous \gls{dmrg} procedure.
Now, let $\lambda = \braket{1|\tilde {\mathcal{L}}|1}$ and define $\alpha$ to minimize 
\begin{align*}
    \argmin_{\alpha} \norm{(\tilde{\mathcal{L}} - \lambda) (\ket{0} + \alpha \ket{1})}.
\end{align*}
This is solved explicitly by defining $Y = \tilde{\mathcal{L}} - \lambda$ and computing 
\begin{align*}
    u_1 &= \braket{0|Y|1} + \braket{1|Y|0}, &
    u_2 &= 2\braket{1|Y|1},
\end{align*}
such that $\alpha = -u_1/u_2$.
We use $\ket{\psi_\alpha} = \ket{0} + \alpha \ket{1}$ as an initial guess for the ground state of nearby $X_{\tilde \lambda}$.
We deem an eigenvector of $X_{\tilde \lambda}$ to be converged when its eigenvalue changes $\epsilon=10^{-4}$.

\section{Different choices for the system operators}\label{app:different_choices}

\begin{figure}
    \centering
    \includegraphics{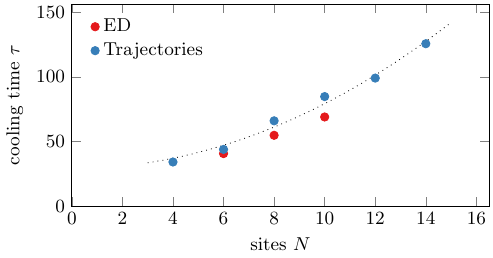}
    \caption{Same as \cref{fig:sshulescaling} but with the system operators in \cref{eq:thermal_system_operators} restricted to $X^\leftarrow$ and $X^\updownarrows_a$. This introduces a directionality into the model while still being consistent with a quadratic fit and thus diffusive dynamics. No error bars shown. }\label{fig:cooling_times_leftonly}
\end{figure}

In \cref{fig:cooling_times_leftonly}, we provide the cooling time in the \gls{ule} when, instead of the full set of operators in \cref{eq:thermal_system_operators} only the system operators $X^\leftarrow$ as well as $X^\updownarrows_a$ are included, thus introducing a large directionality into the model.
This heavily restricted model is still consistent with a quadratic fit and therefore diffusive nature of the cooling.

\section{Mean-field Model of the Engineered Dissipation Dynamics}\label{app:engineered_mean_field}

In this section, we derive the reaction-diffusion equation in \cref{eq:ABfields}.
This is achieved in two steps.
First, we provide a coupled system of equations for the four densities $n_{j,\sigma\sigma'} = x_{j,\sigma}^\dagger x_{j,\sigma'}$.
Then, by introducing a continuum limit, we arrive at the same form of equation as presented in Ref.~\cite{Lyublinskaya_2025}.

We start from the Heisenberg equation of motion for an operator $O$   
\begin{align*}
    \partial_t O=-i[O, H] + \sum_\mu (L_\mu^\dagger O L_\mu - \frac{1}{2} \{L_\mu^\dagger L_\mu, O\}). 
\end{align*}
By explicitly computing the relevant terms (see the Supplemental Material for intermediate steps), we arrive at the following local equations of motion for the four densities:
\begin{align}
    \frac{2}{\kappa} \partial_t n_{k,++} &= -n_{k,++} (5 - 2n_{k,--} + n_{k+1,++} - n_{k+1,--} - n_{k-1,++} - n_{k-1,--} + n_{k-1,+-} + n_{k-1,-+}) \nonumber\\
    &\qquad + n_{k+1,++} + \frac{1}{2} (1 - n_{k+1,--} + n_{k+1,++}) (n_{k,+-} + n_{k,-+}), \label{eq:eom_full_npp}\\
    \frac{2}{\kappa} \partial_t n_{k,--} &= (1 - n_{k,--}) (1 + 2n_{k,++} + n_{k+1,++} + n_{k-1,++} + n_{k-1,--} - n_{k+1,--} - n_{k-1,-+} - n_{k-1,+-}) \nonumber\\
    &\qquad - (1-n_{k+1,--}) + \frac{1}{2} (1-n_{k+1,--}+n_{k+1,++})(n_{k,+-}+n_{k,-+}), \label{eq:eom_full_nmm} \\
    \frac{2}{\kappa} \partial_t n_{k,+-} &= n_{k,+-} (-4-n_{k+1,++} + n_{k+1,--}) + \frac{1}{2} (n_{k,++} + n_{k,--}) (1 - n_{k+1,--} + n_{k+1,++}) - n_{k+1,++},\label{eq:eom_full_npm} \\
    \frac{2}{\kappa} \partial_t n_{k,-+} &= n_{k,-+} (-4-n_{k+1,++} + n_{k+1,--}) + \frac{1}{2} (n_{k,++} + n_{k,--}) (1 - n_{k+1,--} + n_{k+1,++}) - n_{k+1,++}.\label{eq:eom_full_nmp}
\end{align}
All of the above equations have been checked explicitly using computer algebra systems to avoid involuntary errors.

\subsection{Linear Deviation from the Steady State}
\Cref{eq:eom_full_npp,eq:eom_full_nmm,eq:eom_full_npm,eq:eom_full_nmp} can be formulated as field equations to allow for a comparison to previous literature~\cite{Lyublinskaya_2025}. 
Crucially, we find that they do not contain impact ionization terms, and thus do not corroborate the instability scenario.

We use Wick's theorem to turn \cref{eq:eom_full_npp,eq:eom_full_nmm,eq:eom_full_npm,eq:eom_full_nmp} into equations for the expectation values.
We neglect anomalous expectation values of type $\langle x^\dagger x^\dagger\rangle=\langle xx\rangle=0$, as they vanish for the number-conserving state.
Denote $G_{\alpha,\beta}(u,v)=\langle x_{u,\alpha}^\dagger x_{v,\beta}\rangle$ with $\alpha,\beta=\pm$.
Assuming to be in a regime where Wick's theorem holds, i.e., in the late-time limit where the state is approximately Gaussian, we use the identity
\begin{align*}
    \langle n_{u,\alpha\beta} n_{v,\gamma\sigma}\rangle &=\langle x_{u,\alpha}^\dagger x_{u,\beta} x_{v,\gamma}^\dagger x_{v,\sigma}\rangle = \langle x_{u,\alpha}^\dagger x_{u,\beta} \rangle \langle x_{v,\gamma}^\dagger x_{v,\sigma}\rangle + \langle x_{u,\alpha}^\dagger  x_{v,\sigma} \rangle \langle x_{u,\beta}x_{v,\gamma}^\dagger \rangle \\
    &= G_{\alpha\beta}(u,u) G_{\gamma\sigma}(v,v) + G_{\alpha\sigma}(u,v) (\delta_{uv} \delta_{\beta\gamma} -  G_{\gamma\beta}(v,u)).
\end{align*}

From this, we derive field equation by introducing the lattice spacing $a$ and fields $\psi(x=a\xi)=G_{++}(\xi)$, $\Phi(x=a\xi)=1-G_{--}(\xi)$, $\alpha(x=a\xi)=G_{+-}(\xi)$ and $\beta(x=a\xi)=G_{-+}(\xi)$.
We denote spatial derivatives by apostrophes, i.e., $g'=\partial_x g$ and $g''=\partial_x^2 g$, such that the equations of motion for these new fields read
\begin{align*}
    \frac{2}{\kappa} \partial_t \psi &= -4\psi \Phi + \frac{1}{2} (\Phi - \psi) (\alpha+\beta) - 2\alpha\beta + a[\psi(-2\psi'+\alpha'+\beta') + \psi' + \frac{1}{2} (\Phi'+\psi')(\alpha+\beta)] \nonumber\\
    &\qquad + \frac{a^2}{2}[\psi (-2\Phi'' - \alpha'' - \beta'') + \psi'' + \frac{1}{2} (\Phi'' + \psi'')(\alpha+\beta)] + \mathcal{O}(a^3), \\
    \frac{2}{\kappa} \partial_t \Phi &= -4\psi\Phi + \frac{1}{2} (\Phi-\psi)(\alpha+\beta) - 2\alpha\beta +a [\Phi(-2\Phi'-\alpha'-\beta')+\Phi'-\frac{1}{2}(\Phi'+\psi')(\alpha+\beta)] \nonumber\\
    &\qquad +\frac{a^2}{2} [\Phi(-2\psi''+\alpha''+\beta'') + \Phi'' - \frac{1}{2} (\Phi''+\psi'')(\alpha+\beta)] + \mathcal{O}(a^3),\\
    \frac{2}{\kappa} \partial_t \alpha &= \alpha (-3-\psi-\Phi) + \frac{1}{2} (\psi^2-\psi -\Phi^2+\Phi) +a [-\alpha(\psi'+\Phi')+ \frac{1}{2} (\psi+1-\Phi) (\Phi'+\psi') - \psi'] \nonumber \\
    &\qquad +\frac{a^2}{2} [-\alpha(\psi''+\Phi'')+ \frac{1}{2} (\psi+1-\Phi) (\Phi''+\psi'') - \psi''] + \mathcal O(a^3),\\
    \frac{2}{\kappa} \partial_t \beta &= \beta (-3-\psi-\Phi) + \frac{1}{2} (\psi^2-\psi-\Phi^2+\Phi) +a [-\beta(\psi'-\Phi')+ \frac{1}{2} (\psi+1-\Phi) (\Phi'+\psi') - \psi'] \nonumber \\
    &\qquad +\frac{a^2}{2} [-\beta(\psi''+\Phi'')+ \frac{1}{2} (\psi+1-\Phi) (\Phi''+\psi'') - \psi''] + \mathcal O(a^3).
\end{align*}
These may be linearized for small fields $\psi,\Phi,\alpha,\beta\ll 1$.
For this, we assume that their derivatives are sufficiently continuous as well, such that they are on the order of the fields; this should, however, have already been used when we applied Wick's theorem.
The field equations are then
\begin{align}
    \frac{2}{\kappa} \partial_t \psi &= -4\psi \Phi + a\psi' + \frac{a^2}{2}\psi'' + \mathcal{O}(a^3), \\
    \frac{2}{\kappa} \partial_t \Phi &= -4\psi\Phi +a \Phi' +\frac{a^2}{2} \Phi'' + \mathcal{O}(a^3), \\
    \frac{2}{\kappa} \partial_t \alpha &= -3\alpha + \frac{1}{2} (\Phi-\psi) + \frac{a}{2} (\Phi'-\psi') +\frac{a^2}{4} (\Phi''-\psi'') + \mathcal O(a^3) = -3\alpha + \frac{1}{2} e^{a\partial_x}(\Phi-\psi),\\
    \frac{2}{\kappa} \partial_t \beta &= -3\beta + \frac{1}{2} (\Phi-\psi) + \frac{a}{2} (\Phi'-\psi')  +\frac{a^2}{4} (\Phi''-\psi'') + \mathcal O(a^3)  = -3\beta + \frac{1}{2} e^{a\partial_x}(\Phi-\psi).
\end{align}
Note that---in a spatially almost homogeneous case with negligible spatial derivative---the fields $\alpha$ and $\beta$ decay exponentially upon time evolution to their fixed point $\alpha,\beta\approx(\Phi-\psi)/6$.
This will be used later for corrections with $\alpha,\beta\neq 0$.

For now, it is sufficient to note that the fixed point is however much smaller than both the fields $\Phi$ and $\psi$.
This holds especially in the long-time limit, where both fields tend to be the same value such that their difference is small even without accounting for the suppression by the additional factor $1/6$.
Thus, we start with the equations of motion for $\alpha=\beta=0$, which are 
\begin{align*}
    \frac{2}{\kappa} \partial_t \psi &= -4\psi \Phi + a(1-2\psi)\psi' + \frac{a^2}{2}[-2\psi \Phi'' + \psi''] + \mathcal{O}(a^3), \\
    \frac{2}{\kappa} \partial_t \Phi &= -4\psi\Phi +a (1-2\Phi)\Phi' +\frac{a^2}{2} [-2\Phi\psi'' + \Phi''] + \mathcal{O}(a^3),
\end{align*}

The above equations already capture the dynamics contained within this mean-field description. However, previous literature~\cite{Lyublinskaya_2025} formulates them using different quantities. When rewritten, we find that they are absent from the impact ionization terms on which the instability scenario is based.
Concretely, we want to compare this to Ref.~\cite{Lyublinskaya_2025}, by forming the total density deviation $A=(\psi+\Phi)/2$ and the density imbalance $B=(\psi-\Phi)/2$.
In Ref.~\cite[Eq.~(26)]{Lyublinskaya_2025}, they find the \gls{fkpp} equation
\begin{align}
    (\partial_t - D_+ \partial_x^2 - \tau_+^{-1}) A + \frac{\beta}{2} A^2 &= (D_- \partial_x^2 + \tau_-^{-1}) B + \frac{\beta}{2} B^2, \\
    D_- \partial_x^2 A &= (\partial_t - D_+ \partial_x^2) B.
\end{align}
With $\psi=A+B$ and $\Phi=A-B$, our equations look like 
\begin{align}
    \frac{4}{\kappa} \partial_t A &= -8 (A^2 - B^2) + 2a (1-2A)A' - 4aBB' + \frac{a^2}{2} 2(1-2A)A'' + \frac{a^2}{2} 4BB'', \label{eq:AA:1} \\
    \frac{4}{\kappa} \partial_t B &= 2a (1-2A)B' - 4aBA' + \frac{a^2}{2} 2(1+2A)B'' + \frac{a^2}{2} (-4B) A''. \label{eq:AA:2}
\end{align}
To remove the drift terms, we can switch to the comoving reference frame $x' = x - \eta t$ with $4\eta/\kappa=-2a(1-2A)$, such that $\partial_t\rightarrow \partial_t - \eta \partial_x$.
This yields 
\begin{align}
    \Big(\partial_t - \frac{\kappa a^2}{4} (1 -2A) \partial_x^2\Big) A + 2\kappa A^2 &= \Big(-\kappa a B \partial_x + \frac{\kappa a^2}{2} B \partial_x^2 \Big) B + 2\kappa B^2, \label{eq:AB:1} \\
    \Big(-\kappa a B \partial_x - \frac{\kappa a^2}{2} B \partial_x^2  \Big) A &= \Big( \partial_t  - \frac{\kappa a^2}{4} (1 + 2 A) \partial_x^2  \Big) B. \label{eq:AB:2}
\end{align}
By comparing the coefficients, we find 
\begin{align}
    D_+ &= \frac{\kappa a^2}{4} (1 \pm 2A) \xrightarrow{A\rightarrow 0} \frac{\kappa a^2}{4}, &
    D_- &= \pm\frac{\kappa a^2}{2} B\xrightarrow{B\rightarrow 0} 0, &
    \beta &= 4\kappa, &
    v &\equiv \kappa a B \xrightarrow{B\rightarrow 0} 0. \label{eq:coefficients}
\end{align}
The signs in the $D_\pm$ are ill defined far away from the steady state; this issue vanishes for $A,B\rightarrow 0$ close to the steady state. 
In summary, we produced a diffusion equation similar to those presented in Ref.~\cite{Lyublinskaya_2025}, which do not contain the impact ionization term that would render them of \gls{fkpp} type.

\paragraph{Beyond $\alpha=\beta=0$.}

To first order in $a$, we have the steady state value 
\begin{align*}
    \alpha=\beta=\frac{1}{6}(\psi^2-\psi-\Phi^2+\Phi) = \frac{1}{3}(2A-1)B \xrightarrow{A\ll1} -\frac{1}{3}B.
\end{align*}
With this, we get the following additional terms in \cref{eq:AA:1,eq:AA:2}:
\begin{align*}
    \frac{4}{\kappa} \partial_t A &= \text{\cref{eq:AA:1}} - 2B\alpha - 4\alpha^2 + 4aB\alpha' - 2a^2 A \alpha'', \\
    \frac{4}{\kappa} \partial_t B &= \text{\cref{eq:AA:2}} + 4aA\alpha'  - 2a^2 B \alpha'' + 2aA'\alpha + a^2 A'' \alpha.
\end{align*}
Crucially, these are only corrections to the coefficients in \cref{eq:coefficients} and do not change the qualitative behavior of the equations (they also produce quantitatively indistinguishable results).

\subsection{Measuring Impact Ionization}\label{app:impact_ionization}
The destabilizing terms $\propto \tau^{-1} A$ in Ref.~\cite{Lyublinskaya_2025} have been linked to the occurrence of impact ionization.
This impact ionization and its implied instability of the ground state correspond to a process in which an excitation ($x_+$) hops and thus excites an already equilibrated particle; in our model such a process can be mediated by $L_{j,a}^+$.

\begin{figure}
    \centering
    \includegraphics[width=0.45\linewidth]{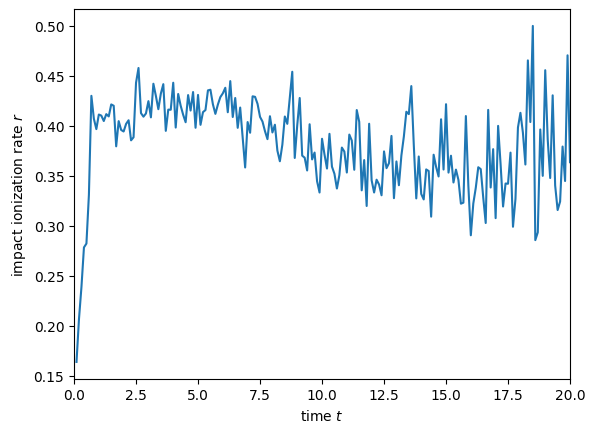}
    \includegraphics[width=0.45\linewidth]{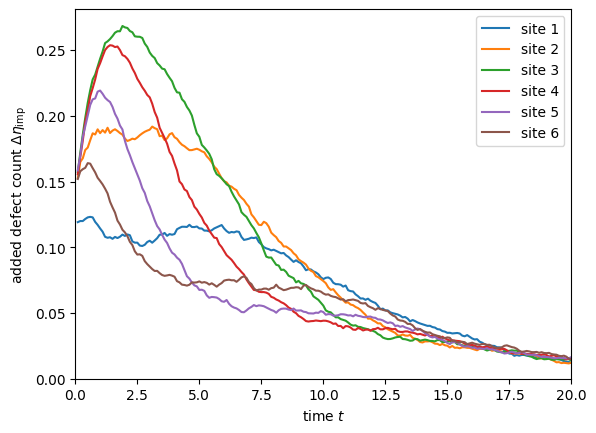}
    \caption{The impact ionization rate $r$ and the added defect count $\Delta \eta_{\mathrm{imp}}$ for time evolution starting from $\ket{\iota_6}$ on an $L=6$ system averaged over $m=10000$ trajectories. While roughly $40\%$ of the jumps are ionizing events, the amount of defects created by it decreases significantly as the steady state is reached, as expected.}
    \label{fig:ABiniota}
\end{figure}

To measure the effect of the impact ionization, we prepare the system in the state $\ket{\iota_n}$, cf.~\cref{eq:iotastate}, which features a field configuration susceptible to impact ionization by $L_{j,a}^+$.
The rate of impact ionization $r$ is measured by counting the number of ionizing jumps, where the dark state occupation after the jump is lower than before the jump, and dividing by the total number of jumps observed in a time period.
We may also define the added defect count $\Delta \eta_{\mathrm{imp}} = \eta(t+\epsilon) - \eta(t)$ created by the ionizing event at time $t$.
In \cref{fig:ABiniota}, we show that while the ionizing events constitute approximately the same number of jump events (which is to be expected, as the nature of the jump operators does not change), their overall effect on the dark state occupation diminishes.
The latter is destroying this apparent instability.

\subsection{Comparing to Numerical Data}
\begin{figure}
    \centering
    \begin{minipage}{0.32\linewidth}
        \centering 
        Exact
    \end{minipage}
    \begin{minipage}{0.32\linewidth}
        \centering 
        \cref{eq:AA:1,eq:AA:2}
    \end{minipage}
    \begin{minipage}{0.32\linewidth}
        \centering 
        Finite size correction, \cref{eq:additional_term}
    \end{minipage}
    \includegraphics[width=0.32\linewidth]{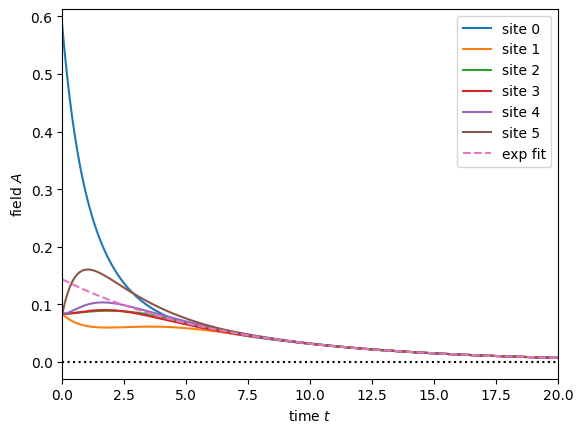}
    \includegraphics[width=0.32\linewidth]{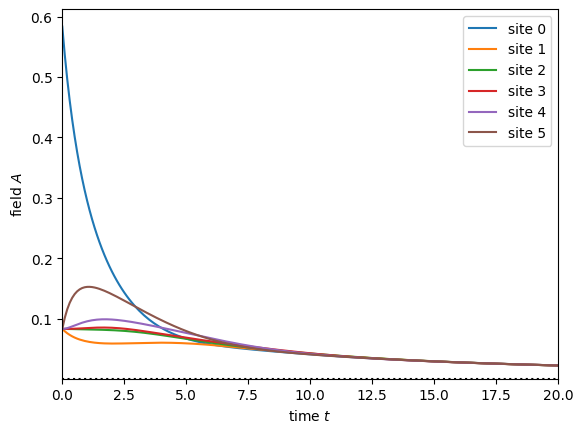}
    \includegraphics[width=0.32\linewidth]{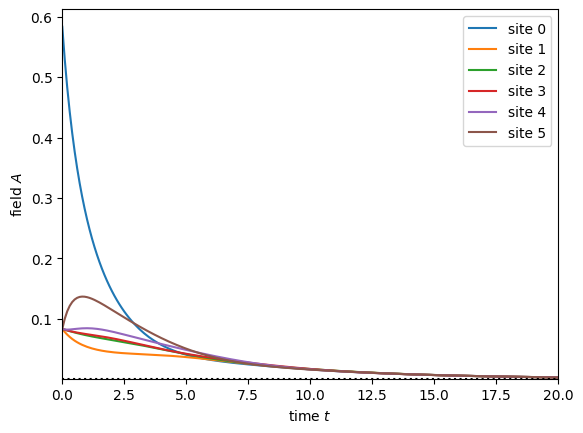}
    \includegraphics[width=0.32\linewidth]{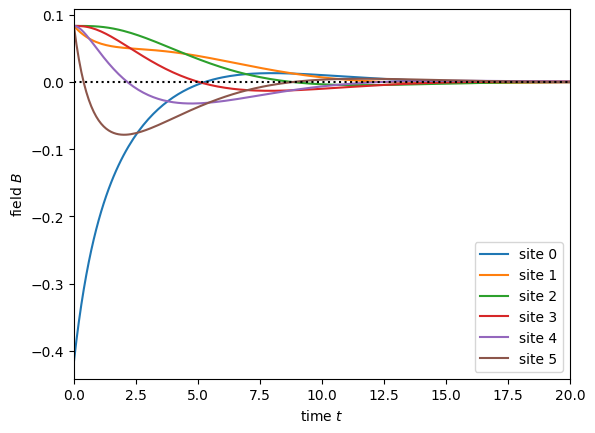}
    \includegraphics[width=0.32\linewidth]{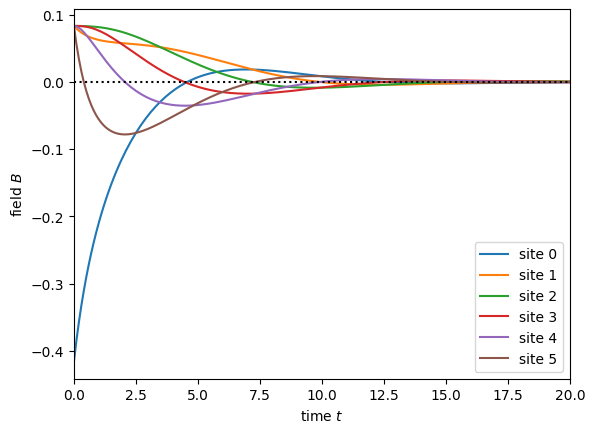}
    \includegraphics[width=0.32\linewidth]{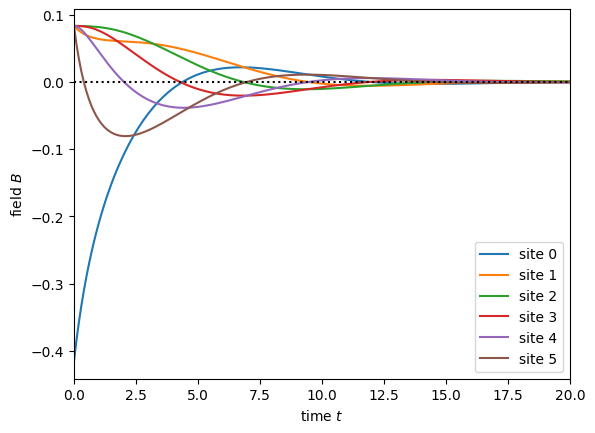}
    \caption{Time evolution (cf.~\cref{fig:ABiniota} for state and parameters) of the fields $A$ and $B$ on different sites. While the field dynamics given in \cref{eq:AA:1,eq:AA:2} roughly capture the short term behavior, for $B\approx 0$ they predict algebraic decay of $A$, which does not match the observed exponential decay. Upon adding the finite-size correction term in \cref{eq:additional_term}, this can be fixed.}
    \label{fig:ABfields}
\end{figure}

\begin{figure}
    \centering
    \includegraphics[width=0.4\linewidth]{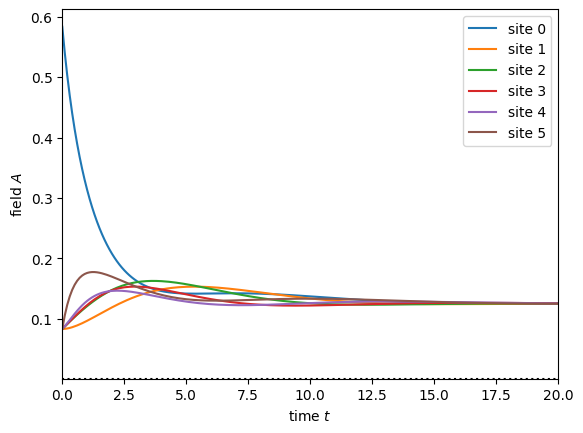}
    \includegraphics[width=0.4\linewidth]{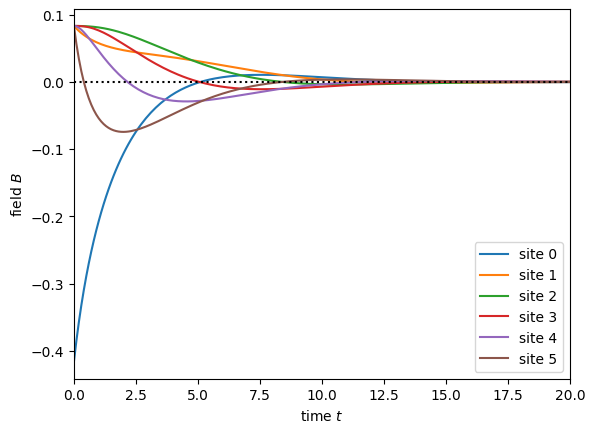}
    \caption{Same as \cref{fig:ABfields} but with an additional non-vanishing impact ionization term $\tau_+ = \tau_-=1$. The field $A$ takes a non-vanishing value for $t \rightarrow \infty$, which can however be ruled out in our model with certainty.}
    \label{fig:AB_with_impact_ionization}
\end{figure}

Finally, we compare the time evolution of the fields $A$ and $B$ to those obtained by implementing \cref{eq:AB:1,eq:AB:2}.
In \cref{fig:ABfields} we show the results of this simulation (second column) against data (first column).
While the time evolutions agree initially, \cref{eq:AB:1,eq:AB:2} dictate algebraic decay in the long-time limit, while we observe exponential decay in the exact numerics.
This is due to the finite size of our system (in this case $L=6$, which is quite small).
This can be accounted for by adding the finite size correction term 
\begin{align}
    \partial_t A = \cdots -\lambda_L A  \label{eq:additional_term}
\end{align}
with $\lambda_L\rightarrow 0$ for $L\rightarrow \infty$ the gap of the system.
It is crucial to note that the new $\lambda_L$ is not the impact ionization term $-\tau_+^{-1}$ of Ref.~\cite{Lyublinskaya_2025}, which differs in sign; the opposite sign would cause an instability.
In \cref{fig:AB_with_impact_ionization}, we show the fields $A$ and $B$ with impact ionization $\tau_+=\tau_-=1$.
Clearly, the equilibrated and non-vanishing value of $A$ is not observed in our model.

\section*{Supplementary material}

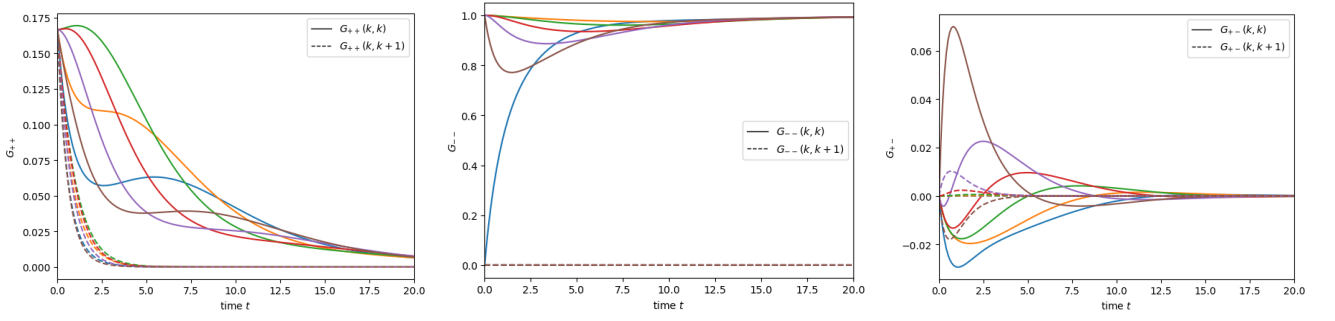

Figure 1. The fields $G_{++}$, $G_{--}$ and $G_{+-}$ evaluated for the state $|\iota_6\rangle$ (introduced in the main text) at different sites (colors). Crucially, for all fields the non-diagonal components (dashed) are small and vanish extremely fast compared to the diagonal components (solid). This motivates neglecting them when taking the continuum limit.

To shorten the equations, we change the notation to $G_{\alpha\beta}(u,v) \to G_{\alpha\beta}^{(u,v)}$ if $u \neq v$ and $G_{\alpha\beta}(u,u) \to G_{\alpha\beta}^{(u)}$ if they are equal. With this, Eqs. (2) to (5) become

$$
\begin{aligned}
\frac{2}{\kappa}\partial_t G_{++}^{(k)} =\ & G_{++}^{(k)}(-5 + 2G_{--}^{(k)} - G_{++}^{(k+1)} + G_{--}^{(k+1)} + G_{++}^{(k-1)} + G_{--}^{(k-1)} - G_{+-}^{(k-1)} - G_{-+}^{(k-1)}) + G_{++}^{(k+1)} \\
& + \frac{1}{2}(1 - G_{--}^{(k+1)} + G_{++}^{(k+1)})(G_{+-}^{(k)} + G_{-+}^{(k)}) - 2G_{+-}^{(k)}G_{-+}^{(k)} \\
& + G_{++}^{(k,k-1)}(G_{-+}^{(k-1,k)} - G_{++}^{(k-1,k)}) + G_{+-}^{(k,k-1)}(G_{++}^{(k-1,k)} - G_{-+}^{(k-1,k)}) \\
& + \frac{1}{2}G_{++}^{(k,k+1)}(G_{++}^{(k+1,k)} - G_{+-}^{(k+1,k)}) + \frac{1}{2}G_{++}^{(k,k+1)}(G_{++}^{(k,k+1)} - G_{-+}^{(k,k+1)}) \\
& + \frac{1}{2}G_{+-}^{(k,k+1)}(G_{--}^{(k+1,k)} - G_{-+}^{(k+1,k)}) + \frac{1}{2}G_{-+}^{(k,k+1)}(G_{--}^{(k,k+1)} - G_{+-}^{(k,k+1)}),
\end{aligned}
\tag{6}
$$

$$
\begin{aligned}
\frac{2}{\kappa}\partial_t G_{--}^{(k)} =\ & (1 - G_{--}^{(k)})(1 + 2G_{++}^{(k)} + G_{++}^{(k+1)} + G_{++}^{(k-1)} + G_{--}^{(k-1)} - G_{--}^{(k-1)} - G_{-+}^{(k-1)} - G_{+-}^{(k-1)}) - (1 - G_{--}^{(k+1)}) \\
& + \frac{1}{2}(1 - G_{--}^{(k+1)} + G_{++}^{(k+1)})(G_{+-}^{(k)} + G_{-+}^{(k)}) + 2G_{+-}^{(k)}G_{-+}^{(k)} \\
& - G_{-+}^{(k,k-1)}(G_{--}^{(k-1,k)} - G_{+-}^{(k-1,k)}) - G_{--}^{(k,k-1)}(G_{+-}^{(k-1,k)} - G_{--}^{(k-1,k)}) \\
& - \frac{1}{2}G_{--}^{(k,k+1)}(G_{--}^{(k+1,k)} - G_{-+}^{(k+1,k)}) - \frac{1}{2}G_{--}^{(k,k+1)}(G_{--}^{(k,k+1)} - G_{+-}^{(k,k+1)}) \\
& - \frac{1}{2}G_{-+}^{(k,k+1)}(G_{++}^{(k+1,k)} - G_{+-}^{(k+1,k)}) - \frac{1}{2}G_{+-}^{(k,k+1)}(G_{++}^{(k,k+1)} - G_{-+}^{(k,k+1)}).
\end{aligned}
\tag{7}
$$

Note that the total fermion number is still conserved, i.e., $\partial_t(\sum_k G_{++} + G_{--}) = 0$. For the correlators, we obtain

$$
\begin{aligned}
\frac{2}{\kappa}\partial_t G_{+-}^{(k)} =\ & G_{+-}^{(k)}(-4 - G_{++}^{(k+1)} + G_{--}^{(k+1)}) + \frac{1}{2}(G_{++}^{(k)} + G_{--}^{(k)})(1 - G_{--}^{(k+1)} + G_{++}^{(k+1)}) - G_{++}^{(k+1)} \\
& + \frac{1}{2}G_{++}^{(k,k+1)}(G_{+-}^{(k+1,k)} - G_{++}^{(k+1,k)}) + \frac{1}{2}G_{+-}^{(k+1,k)}(G_{++}^{(k,k+1)} - G_{-+}^{(k+1,k)}) \\
& + \frac{1}{2}G_{+-}^{(k,k+1)}(G_{-+}^{(k+1,k)} - G_{--}^{(k+1,k)}) + \frac{1}{2}G_{+-}^{(k+1,k)}(G_{--}^{(k,k+1)} - G_{+-}^{(k,k+1)}),
\end{aligned}
\tag{8}
$$

$$
\begin{aligned}
\frac{2}{\kappa}\partial_t G_{-+}^{(k)} =\ & G_{-+}^{(k)}(-4 - G_{++}^{(k+1)} + G_{--}^{(k+1)}) + \frac{1}{2}(G_{++}^{(k)} + G_{--}^{(k)})(1 - G_{--}^{(k+1)} + G_{++}^{(k+1)}) - G_{++}^{(k+1)} \\
& + \frac{1}{2}G_{-+}^{(k,k+1)}(G_{++}^{(k+1,k)} - G_{+-}^{(k,k+1)}) + \frac{1}{2}G_{++}^{(k+1,k)}(G_{-+}^{(k,k+1)} - G_{++}^{(k,k+1)}) \\
& + \frac{1}{2}G_{-+}^{(k+1,k)}(G_{+-}^{(k,k+1)} - G_{--}^{(k,k+1)}) + \frac{1}{2}G_{--}^{(k,k+1)}(G_{--}^{(k+1,k)} - G_{-+}^{(k+1,k)}).
\end{aligned}
\tag{9}
$$

## B. Towards a continuum limit

Equations (6) to (9) are defined on a discrete lattice and contain the $G$ as a function of two parameters. To move forward, we ignore all non-diagonal contributions $G(x, y)$ with $x \neq y$. This is motivated by comparing with numerical results in Fig. 1, where neglecting all non-diagonal contributions provides an accurate description.

Next, we introduce the lattice spacing $a$ and fields $\psi(x = a\xi) = G_{++}(\xi)$, $\phi(x = a\xi) = G_{--}(\xi)$, $\alpha(x = a\xi) = G_{+-}(\xi)$ and $\beta(x = a\xi) = G_{-+}(\xi)$. We denote derivatives by apostrophes, i.e., $g' = \partial_x g$ and $g'' = \partial_x^2 g$. Then, we obtain

$$
\begin{aligned}
\frac{2}{\kappa}\partial_t \psi = {} & a^0[-4\psi(1-\phi) + \frac{1}{2}(1-\phi-\psi)(\alpha+\beta) - 2\alpha\beta] + a^1[\psi(-2\psi'+\alpha'+\beta') + \psi' + \frac{1}{2}(-\phi'+\psi')(\alpha+\beta)] \\
& + \frac{a^2}{2}[\psi(2\phi''-\alpha''-\beta'') + \psi'' + \frac{1}{2}(-\phi''+\psi'')(\alpha+\beta)] + \mathcal{O}(a^3),
\end{aligned}
\tag{10}
$$

$$
\begin{aligned}
\frac{2}{\kappa}\partial_t \phi = {} & a^0[4\psi(1-\phi) - \frac{1}{2}(1-\phi-\psi)(\alpha+\beta) + 2\alpha\beta] + a^1[(1-\phi)(-2\phi'+\alpha'+\beta') + \phi' + \frac{1}{2}(-\phi'+\psi')(\alpha+\beta)] \\
& + \frac{a^2}{2}[(1-\phi)(2\psi''-\alpha''-\beta'') + \phi'' + \frac{1}{2}(-\phi''+\psi'')(\alpha+\beta)] + \mathcal{O}(a^3).
\end{aligned}
\tag{11}
$$

We can take their sum, finding

$$
\begin{aligned}
\frac{2}{\kappa}\partial_t(\psi+\phi) = {} & a^1\partial_x[\psi - \psi^2 - \phi + \phi^2 + (\psi+1-\phi)(\alpha+\beta)] \\
& + \frac{a^2}{2}\partial_x[2\phi'\psi + 2(1-\phi)\psi' + \psi' + \phi' - \alpha' - \beta' + (\phi-\psi)(\alpha'+\beta') - (\phi'-\psi')(\alpha+\beta)] + \mathcal{O}(a^3).
\end{aligned}
$$

Since this is a total derivative, the total fermion number is conserved upon integration $\partial_t \int \mathrm{d}x(\psi+\phi) = 0$ over a domain with periodic boundary conditions. Similarly, we obtain for the correlator fields

$$
\begin{aligned}
\frac{2}{\kappa}\partial_t \alpha = {} & a^0[\alpha(-4-\psi+\phi) + \frac{1}{2}(\psi^2-\phi^2+\phi-\psi)] + a^1[\alpha(-\psi'+\phi') + \frac{1}{2}(\psi+\phi)(-\phi'+\psi') - \psi'] \\
& + \frac{a^2}{2}[\alpha(-\psi''+\phi'') + \frac{1}{2}(\psi+\phi)(-\phi''+\psi'') - \psi''] + \mathcal{O}(a^3),
\end{aligned}
\tag{12}
$$

$$
\begin{aligned}
\frac{2}{\kappa}\partial_t \beta = {} & a^0[\beta(-4-\psi+\phi) + \frac{1}{2}(\psi^2-\phi^2+\phi-\psi)] + a^1[\beta(-\psi'+\phi') + \frac{1}{2}(\psi+\phi)(-\phi'+\psi') - \psi'] \\
& + \frac{a^2}{2}[\beta(-\psi''+\phi'') + \frac{1}{2}(\psi+\phi)(-\phi''+\psi'') - \psi''] + \mathcal{O}(a^3).
\end{aligned}
\tag{13}
$$

To obtain the field equations in the main text, it remains to substitute $\Phi = 1 - \phi$.

## II. THE ENGINEERED MODEL IN THE TRUNCATED HILBERT SPACE

For the engineered dissipation, we find the first order matrix elements of the Lindbladian from the excited states $|p, q\rangle = x_{p,+}^\dagger x_{q,-} |\circ\rangle$ to be $\langle\circ|L\circ\rangle = 0$ and $\langle p, q|L\circ\rangle = 0$ independent of $L$ and $p, q$. For the other matrix elements, define $\alpha_{i,j} = \{x_{\sigma,i}^\dagger, x_{\sigma,j}\}$ the anticommutator, such that we find

$$
\langle\circ|L_{j,a}^+|p, q\rangle = -\sum_{s=0}^{L-1} \lambda^s \alpha_{j-1-s,q} \alpha_{p,j}, \qquad\qquad \langle\circ|L_{j,a}^-|p, q\rangle = -\sum_{s=0}^{L-1} \lambda^s \alpha_{j-1-s,p} \alpha_{q,j},
$$

$$
\langle\circ|L_{j,b}^+|p, q\rangle = +\sum_{s=0}^{L-1} \lambda^s \alpha_{j+s,q} \alpha_{p,j}, \qquad\qquad \langle\circ|L_{j,b}^-|p, q\rangle = -\sum_{s=0}^{L-1} \lambda^s \alpha_{j+s,p} \alpha_{q,j},
$$

for the transition to the dark state. The intraband transitions are given by

$$
\langle p_1, q_1|L_{j,a}^+|p_2, q_2\rangle = +\sum_{s=0}^{L-1} \lambda^s \alpha_{p_1,j-1-s} \alpha_{q_1,q_2} \alpha_{p_2,j}, \qquad \langle p_1, q_1|L_{j,a}^-|p_2, q_2\rangle = -\sum_{s=0}^{L-1} \lambda^s \alpha_{q_1,j-1-s} \alpha_{p_1,p_2} \alpha_{q_2,j},
$$

$$
\langle p_1, q_1|L_{j,b}^+|p_2, q_2\rangle = +\sum_{s=0}^{L-1} \lambda^s \alpha_{p_1,j+s} \alpha_{q_1,q_2} \alpha_{p_2,j}, \qquad \langle p_1, q_1|L_{j,b}^-|p_2, q_2\rangle = +\sum_{s=0}^{L-1} \lambda^s \alpha_{q_1,j+s} \alpha_{p_1,p_2} \alpha_{q_2,j}.
$$

For the second order states $|p_1 p_2, q_1 q_2\rangle = x^\dagger_{p_1+} x^\dagger_{p_2+} x_{q_1-} x_{q_2-} |\circ\rangle$ with $p_1 < p_2$ and $q_1 < q_2$ we find that $\langle\circ|L|pp, qq\rangle = 0$ for the transition to the ground state. For transitions to the first order states, we have

$$\langle p, q|L^+_{j,a}|p_1 p_2, q_1 q_2\rangle = \sum_s \lambda^s (\alpha_{p,p_2}\alpha_{j,p_1} - \alpha_{p,p_1}\alpha_{j,p_2})(\alpha_{j-s-1,q_1}\alpha_{q,q_2} - \alpha_{q,q_1}\alpha_{j-s-1,q_2}),$$

$$\langle p, q|L^-_{j,a}|p_1 p_2, q_1 q_2\rangle = \sum_s \lambda^s (\alpha_{q,q_2}\alpha_{j,q_1} - \alpha_{q,q_1}\alpha_{j,q_2})(\alpha_{j-s-1,p_1}\alpha_{p,p_2} - \alpha_{p,p_1}\alpha_{j-s-1,p_2}),$$

$$\langle p, q|L^+_{j,b}|p_1 p_2, q_1 q_2\rangle = \sum_s \lambda^s (\alpha_{j,p_2}\alpha_{p,p_1} - \alpha_{j,p_1}\alpha_{p,p_2})(\alpha_{j+s,q_1}\alpha_{q,q_2} - \alpha_{q,q_1}\alpha_{j+s,q_2}),$$

$$\langle p, q|L^-_{j,b}|p_1 p_2, q_1 q_2\rangle = \sum_s \lambda^s (\alpha_{j,q_1}\alpha_{q,q_2} - \alpha_{j,q_2}\alpha_{q,q_1})(\alpha_{j+s,p_1}\alpha_{p,p_2} - \alpha_{p,p_1}\alpha_{j+s,p_2}).$$

Now, at last, for the intraband transitions in second order

$$\langle p_1 p_2, q_1 q_2|L^+_{ja}|p_3 p_4, q_3 q_4\rangle = + \sum_s \lambda^s [\alpha_{j,p_3}(\alpha_{p_1,j-s-1}\alpha_{p_4,p_2} - \alpha_{p_1,p_4}\alpha_{j-s-1,p_2}) - \alpha_{j,p_4}(\alpha_{p_1,j-s-1}\alpha_{p_2,p_3} - \alpha_{p_1,p_3}\alpha_{p_2,j-s-1})]$$
$$\times (\alpha_{q_1,q_3}\alpha_{q_2,q_4} - \alpha_{q_2,q_3}\alpha_{q_1,q_4})$$

$$\langle p_1 p_2, q_1 q_2|L^-_{ja}|p_3 p_4, q_3 q_4\rangle = - \sum_s \lambda^s [\alpha_{j,q_3}(\alpha_{q_1,j-s-1}\alpha_{q_4,q_2} - \alpha_{q_1,q_4}\alpha_{j-s-1,q_2}) - \alpha_{j,q_4}(\alpha_{q_1,j-s-1}\alpha_{q_3,q_2} - \alpha_{q_1,q_3}\alpha_{j-s-1,q_2})]$$
$$\times (\alpha_{p_1,p_3}\alpha_{p_2,p_4} - \alpha_{p_2,p_3}\alpha_{p_1,p_4})$$

$$\langle p_1 p_2, q_1 q_2|L^+_{jb}|p_3 p_4, q_3 q_4\rangle = + \sum_s \lambda^s [\alpha_{j,p_3}(\alpha_{p_1,j+s}\alpha_{p_2,p_4} - \alpha_{p_1,p_4}\alpha_{p_2,j+s}) - \alpha_{j,p_4}(\alpha_{p_1,j+s}\alpha_{p_2,p_3} - \alpha_{p_1,p_3}\alpha_{p_2,j+s})]$$
$$\times (\alpha_{q_1,q_3}\alpha_{q_2,q_4} - \alpha_{q_1,q_4}\alpha_{q_2,q_3})$$

$$\langle p_1 p_2, q_1 q_2|L^-_{jb}|p_3 p_4, q_3 q_4\rangle = + \sum_s \lambda^s [\alpha_{j,q_3}(\alpha_{q_1,j+s}\alpha_{q_2,q_4} - \alpha_{q_1,q_4}\alpha_{q_2,j+s}) - \alpha_{j,q_4}(\alpha_{q_1,j+s}\alpha_{q_2,q_3} - \alpha_{q_1,q_3}\alpha_{q_2,j+s})]$$
$$\times (\alpha_{p_1,p_3}\alpha_{p_2,p_4} - \alpha_{p_1,p_4}\alpha_{p_2,p_3}).$$

## III. THE THERMAL MODEL IN THE TRUNCATED HILBERT SPACE

The system-bath interaction is described by $M$ operators $X_\alpha$ acting on the system, which model the change induced by the bath. The universal Lindblad equation for spectral density $J$ is then given by the with $M$ Jump operators

$$L_\lambda = \sum_\alpha \sum_{nm} \sqrt{J_{\lambda\alpha}(E_n - E_m)} |m\rangle\langle m| X_\alpha |n\rangle\langle n|,$$

where $|m\rangle$, $|n\rangle$ are eigenstates of the system Hamiltonian. As we consider system at very low temperatures, i.e., $\beta \approx 10$, the spectral density suppresses all jumps leaving the ground state. The spectral density is chosen to be $J_{\lambda\alpha} = \delta_{\lambda\alpha} J$ with

$$J(\omega; \beta) = \frac{1}{2 + \ln(1 + \beta\Lambda_0)} \frac{e^{-\omega^2/2\Lambda_0^2}}{1 + e^{-\beta\omega}}.$$

For two frequencies $\omega_1$ and $\omega_2$, we have

$$\frac{J(\omega_1)}{J(\omega_2)} = e^{-(\omega_1^2 - \omega_2^2)/\Lambda_0^2} \frac{1 + e^{-\beta\omega_2}}{1 + e^{-\beta\omega_1}}.$$

We consider the $M = 6L$ jump operators given by nearest-neighbor hopping

$$X^1_\alpha = |n, X\rangle\langle n + 1, X|, \qquad X^2_\alpha = |n + 1, X\rangle\langle n, X|, \qquad X^3_\alpha = |n, X\rangle\langle n, 1 - X|.$$

From now on, we switch to creation and annihilation operators of the eigenstates of the dimerized Su–Schrieffer–Heeger (SSH) Hamiltonian in the manuscript. The Hamiltonian is readily diagonalized by introducing the Fourier transformed operators $c^\dagger_{k,\gamma} = \frac{1}{\sqrt{L}} \sum_j e^{-ikj} c^\dagger_{j,\gamma}$ with lattice momentum $k = 2\pi m/L$ ($0 \le m < L$), such that the Hamiltonian is almost diagonal with

$$H = \sum_k \boldsymbol{c}^\dagger_k \begin{pmatrix} 0 & f(k) \\ f^*(k) & 0 \end{pmatrix} \boldsymbol{c}_k,$$

with $\boldsymbol{c}_k = (c_{a,k}, c_{b,k})^{\mathrm{T}}$ and $f(k) = te^{-ik}$. This system has the dispersion relation

$$E^2(k) = t^2.$$

Furthermore, let $f(k) = r_k e^{i\phi_k}$, then the eigenvectors are created by

$$c^\dagger_{k,\pm} = \frac{1}{\sqrt{2}}(\pm e^{i\phi_k} c^\dagger_{k,a} + c^\dagger_{k,b}).$$

Explicitly, the phase is given by

$$\tan\phi_k = \frac{-t\sin k}{t\cos k} = -\tan k.$$

Lastly, let

$$c^\dagger_{kA} = \frac{e^{-i\phi_k}}{\sqrt{2}}(c^\dagger_{k+} - c^\dagger_{k-}), \qquad c^\dagger_{kB} = \frac{1}{\sqrt{2}}(c^\dagger_{k+} + c^\dagger_{k-}), \qquad c^\dagger_{nX} = \frac{1}{\sqrt{2L}}\sum_k e^{-ikn - iX\phi_k}(c^\dagger_{k+} + (-1)^X c^\dagger_{k-}).$$

Now, we set $\chi = 1$ ($\chi = 0$) for subsystems $A$ ($B$) respectively. We may now explicitly state the jump operators in the eigenbasis of the SSH Hamiltonian

$$\begin{aligned}
X^1_{n\chi} = c^\dagger_{n,\chi} c_{n+1,\chi} &= \frac{1}{2L}\sum_{kk'} e^{-ikn - i\chi\phi_k}(c^\dagger_{k+} + (-1)^\chi c^\dagger_{k-})e^{ik'(n+1) + i\chi\phi_{k'}}(c_{k'+} + (-1)^\chi c_{k'-}) \\
&= \frac{1}{2L}\sum_{kk'} e^{-i(k-k')n + ik' - iX(\phi_k - \phi_{k'})}(c^\dagger_{k+} + (-1)^\chi c^\dagger_{k-})(c_{k'+} + (-1)^\chi c_{k'-}) \\
&\equiv \frac{1}{2L}\sum_{kk'} \underbrace{e^{-i(k-k')n + ik' - iX(\phi_k - \phi_{k'})}}_{a^1(k,k',X)} Q_{kk'\chi},
\end{aligned}$$

for the first kind as well as

$$\begin{aligned}
X^2_{n\chi} = c^\dagger_{n+1,\chi} c_{n,\chi} &= \frac{1}{2L}\sum_{kk'} e^{-ik(n+1) - i\chi\phi_k}(c^\dagger_{k+} + (-1)^\chi c^\dagger_{k-})e^{ik'n + i\chi\phi_{k'}}(c_{k'+} + (-1)^\chi c_{k'-}) \\
&= \frac{1}{2L}\sum_{kk'} \underbrace{e^{-i(k-k')n - ik - i\chi(\phi_k - \phi_{k'})}}_{a^2(k,k',\chi)} Q_{kk'\chi},
\end{aligned}$$

for the second kind and lastly

$$\begin{aligned}
X^3_{n\chi} = c^\dagger_{n,\chi} c_{n,1-\chi} &= \frac{1}{2L}\sum_{kk'} e^{-ikn - i\chi\phi_k}(c^\dagger_{k+} + (-1)^\chi c^\dagger_{k-})e^{ik'n + i(1-\chi)\phi_{k'}}(c_{k'+} + (-1)^{1-\chi} c_{k'-}) \\
&= \frac{1}{2L}\sum_{kk'} \underbrace{e^{-i(k-k')n - i\chi(\phi_k - \phi_{k'}) + i\phi_k}}_{a^3(k,k',\chi)} \hat{Q}_{kk'\chi}.
\end{aligned}$$

Notice that the while the prefactors vary for the three kinds of system operators $X$, the operators facilitating the cooling are shared. Therefore, we consider the matrix elements of $Q_{kk'\chi} = (c^\dagger_{k+} + (-1)^\chi c^\dagger_{k-})(c_{k'+} + (-1)^\chi c_{k'-})$. For this, define the truncated Hilbert space consisting of the ground state and one-particle excitations

$$|g\rangle = \prod_k c^\dagger_{k-} |0\rangle, \qquad\qquad\qquad |ij\rangle = c^\dagger_{k_i+} c_{k_j-} |g\rangle.$$

We find the matrix elements of $Q$ in this truncated basis to be

$$\begin{aligned}
\langle g|\, Q_{kk'\chi}\, |g\rangle &= \langle g|\, (c^\dagger_{k+} + (-1)^\chi c^\dagger_{k-})(c_{k'+} + (-1)^\chi c_{k'-})\, |g\rangle = \delta_{kk'}, \\
\langle g|\, Q_{kk'\chi}\, |ij\rangle &= \langle g|\, (c^\dagger_{k+} + (-1)^\chi c^\dagger_{k-})(c_{k'+} + (-1)^\chi c_{k'-}) c^\dagger_{k_i+} c_{k_j-} |g\rangle = (-1)^\chi \delta_{k'k_i}\delta_{kk_j}, \\
\langle ij|\, Q_{kk'\chi}\, |g\rangle &= \langle g|\, c^\dagger_{k_j-} c_{k_i+}(c^\dagger_{k+} + (-1)^\chi c^\dagger_{k-})(c_{k'+} + (-1)^\chi c_{k'-}) |g\rangle = (-1)^\chi \delta_{kk_i}\delta_{k'k_j}, \\
\langle ij|\, Q_{kk'\chi}\, |i'j'\rangle &= \langle g|\, c^\dagger_{k_j-} c_{k_i+}(c^\dagger_{k+} + (-1)^\chi c^\dagger_{k-})(c_{k'+} + (-1)^\chi c_{k'-}) c^\dagger_{k_{i'}+} c_{k_{j'}-} |g\rangle \\
&= \delta_{k_j k_{j'}}\delta_{k'k_{i'}}\delta_{kk_i} - \delta_{k_i k_{i'}}(\delta_{kk_j}\delta_{k'k_j} - \delta_{kk'}\delta_{k_j k_{j'}}).
\end{aligned}$$

Similarly, for $\hat{Q}_{kk'\chi} = (c_{k+}^\dagger + (-1)^\chi c_{k-}^\dagger)(c_{k'+} + (-1)^{1-\chi} c_{k'-})$ one obtains the matrix elements

$$\langle g | \hat{Q}_{kk'\chi} | g \rangle = \langle g | (c_{k+}^\dagger + (-1)^\chi c_{k-}^\dagger)(c_{k'+} + (-1)^{1-\chi} c_{k'-}) | g \rangle = -\delta_{kk'},$$

$$\langle g | \hat{Q}_{kk'\chi} | ij \rangle = \langle g | (c_{k+}^\dagger + (-1)^\chi c_{k-}^\dagger)(c_{k'+} + (-1)^{1-\chi} c_{k'-}) c_{i+}^\dagger c_{k_j-} | g \rangle = (-1)^\chi \delta_{k'k_i} \delta_{kk_j},$$

$$\langle ij | \hat{Q}_{kk'\chi} | g \rangle = \langle g | c_{k_j-}^\dagger c_{k_i+} (c_{k+}^\dagger + (-1)^\chi c_{k-}^\dagger)(c_{k'+} + (-1)^{1-\chi} c_{k'-}) | g \rangle = (-1)^{1-\chi} \delta_{kk_i} \delta_{k'k_j},$$

$$\langle ij | \hat{Q}_{kk'\chi} | i'j' \rangle = \langle g | c_{k_j-}^\dagger c_{k_i+} (c_{k+}^\dagger + (-1)^\chi c_{k-}^\dagger)(c_{k'+} + (-1)^{1-\chi} c_{k'-}) c_{k_{i'}+}^\dagger c_{k_{j'}-} | g \rangle$$
$$= \delta_{k_j k_{j'}} \delta_{k'k_{i'}} \delta_{kk_i} + \delta_{k_i k_{i'}} (\delta_{kk_{j'}} \delta_{k'k_j} - \delta_{kk'} \delta_{k_j k_{j'}}).$$

Now for the fun part. We have found that

$$A_{n\chi}^q = \frac{1}{2L} \sum_{kk'} a^q(k, k', \chi) Q_{kk'\chi}.$$

For the density matrix $\rho$, we make the ansatz

$$\rho = W_{ij;kl} | ij \rangle \langle kl | + (1 - W_{ij;ij}) | g \rangle \langle g |$$

and solve the Lindblad equation

$$\dot{\rho} = \mathcal{L}[\rho] = -i[H, \rho] + \sum_\lambda (2L_\lambda \rho L_\lambda^\dagger - \{L_\lambda^\dagger L_\lambda, \rho\}).$$

We are interested in the flat-band case. Thus, let $\Delta$ be the energy of one excitation and define the natural decay rates

$$\gamma_0 = J(0; \beta), \qquad\qquad \gamma_\pm = J(\pm\Delta; \beta).$$

Due to the low temperature $\beta = 10$, we approximate $\gamma_- \approx 0$. We find the matrix elements of the jump operators to be

$$\langle g | L_{n\chi}^q | g \rangle = \frac{\sqrt{\gamma_0}}{2L} \sum_k a^q(k, k, \chi) = \frac{\sqrt{\gamma_0}}{2} \Xi_q,$$

$$\langle g | L_{n\chi}^q | ij \rangle = \frac{\sqrt{\gamma_+}}{2L} (-1)^\chi a^q(k_j, k_i, \chi),$$

$$\langle ij | L_{n\chi}^q | g \rangle = \frac{\sqrt{\gamma_-}}{2L} (-1)^\chi a^q(k_i, k_j, \chi) \approx 0,$$

$$\langle ij | L_{n\chi}^q | i'j' \rangle = \frac{\sqrt{\gamma_0}}{2L} \Big[ \delta_{k_j k_{j'}} a^q(k_i, k_{i'}, \chi) - \delta_{k_i k_{i'}} (a^q(k_{j'}, k_j, \chi) - \delta_{k_j k_{j'}} \sum_k a^q(k, k, \chi)) \Big].$$

We may explicitly evaluate $\sum_k a^q(k, k, \chi) = L \Xi_q$, with $\Xi_q = \delta_{q3}$ in the dimerized limit.

## IV. THE THERMAL LINDBLADIAN IN THE BOND BASIS

Here, we provide the explicit eigenspace representation of all system operators $X$ used in the thermal Lindbladian in the bond basis $x_{j,\sigma}$. In this representation, the six system operators on a unit cell in this basis have the representation

$$X_{i,a}^\rightarrow = \frac{1}{2}(x_{i-1,+}^\dagger - x_{i-1,-}^\dagger)(x_{i,+} - x_{i,-}) \qquad\qquad X_{i,a}^\leftarrow = \frac{1}{2}(x_{i-1,+}^\dagger - x_{i-1,-}^\dagger)(x_{i-2,+} - x_{i-2,-})$$

$$X_{i,b}^\rightarrow = \frac{1}{2}(x_{i,+}^\dagger + x_{i,-}^\dagger)(x_{i+1,+} + x_{i+1,-}) \qquad\qquad X_{i,b}^\leftarrow = \frac{1}{2}(x_{i,+}^\dagger + x_{i,-}^\dagger)(x_{i-1,+} + x_{i-1,-})$$

$$X_{i,a}^\updownarrow = \frac{1}{2}(x_{i,+}^\dagger + x_{i,-}^\dagger)(x_{i-1,+} - x_{i-1,-}) \qquad\qquad X_{i,b}^\updownarrow = \frac{1}{2}(x_{i-1,+}^\dagger - x_{i-1,-}^\dagger)(x_{i,+} + x_{i,-})$$

To get to the Lindblad operators, we isolate the individual energy contributions. We give them here separated, as they would be in the Davies formulation.

$$L_{i,a}^{\to 0} = \frac{\sqrt{\gamma(0)}}{2}(x_{i-1,+}^\dagger x_{i,+} + x_{i-1,-}^\dagger x_{i,-}), \qquad L_{i,a}^{\to -} = -\frac{\sqrt{\gamma(2)}}{2}(x_{i-1,-}^\dagger x_{i,+}), \qquad L_{i,a}^{\to +} = -\frac{\sqrt{\gamma(-2)}}{2}(x_{i-1,+}^\dagger x_{i,-}),$$

$$L_{i,b}^{\to 0} = \frac{\sqrt{\gamma(0)}}{2}(x_{i,+}^\dagger x_{i+1,+} + x_{i,-}^\dagger x_{i+1,-}), \qquad L_{i,b}^{\to -} = \frac{\sqrt{\gamma(2)}}{2}(x_{i,-}^\dagger x_{i+1,+}), \qquad L_{i,b}^{\to +} = \frac{\sqrt{\gamma(-2)}}{2}(x_{i,+}^\dagger x_{i+1,-}),$$

$$L_{i,a}^{\leftarrow 0} = \frac{\sqrt{\gamma(0)}}{2}(x_{i-1,+}^\dagger x_{i-2,+} + x_{i-1,-}^\dagger x_{i-2,-}), \qquad L_{i,a}^{\leftarrow -} = -\frac{\sqrt{\gamma(2)}}{2}(x_{i-1,-}^\dagger x_{i-2,+}), \qquad L_{i,a}^{\leftarrow +} = -\frac{\sqrt{\gamma(-2)}}{2}(x_{i-1,+}^\dagger x_{i-2,-}),$$

$$L_{i,b}^{\leftarrow 0} = \frac{\sqrt{\gamma(0)}}{2}(x_{i,+}^\dagger x_{i-1,+} + x_{i,-}^\dagger x_{i-1,-}), \qquad L_{i,b}^{\leftarrow -} = \frac{\sqrt{\gamma(2)}}{2}(x_{i,-}^\dagger x_{i-1,+}), \qquad L_{i,b}^{\leftarrow +} = \frac{\sqrt{\gamma(-2)}}{2}(x_{i,+}^\dagger x_{i-1,-}),$$

$$L_{i,a}^{\updownarrow 0} = \frac{\sqrt{\gamma(0)}}{2}(x_{i,+}^\dagger x_{i-1,+} - x_{i,-}^\dagger x_{i-1,-}), \qquad L_{i,a}^{\updownarrow -} = \frac{\sqrt{\gamma(2)}}{2}(x_{i,-}^\dagger x_{i-1,+}), \qquad L_{i,a}^{\updownarrow +} = -\frac{\sqrt{\gamma(-2)}}{2}(x_{i,+}^\dagger x_{i-1,-}),$$

$$L_{i,b}^{\updownarrow 0} = \frac{\sqrt{\gamma(0)}}{2}(x_{i-1,+}^\dagger x_{i,+} - x_{i-1,-}^\dagger x_{i,-}), \qquad L_{i,b}^{\updownarrow -} = -\frac{\sqrt{\gamma(2)}}{2}(x_{i-1,-}^\dagger x_{i,+}), \qquad L_{i,b}^{\updownarrow +} = \frac{\sqrt{\gamma(-2)}}{2}(x_{i-1,+}^\dagger x_{i,-}).$$

In the low-temperature limit $\beta \Lambda_0 \gg 1$, the contributions involving an energy gain are exponentially surpressed, i.e., $\gamma(-2) \to 0$.

To compare the action of the thermal Lindbladian to the random walk implemented by the engineered operators, we also explicitly compute terms in the effective Hamiltonian $\tilde{H} = H - \frac{i}{2}\sum_\mu A_\mu$, some of which are

$$A_{i,a}^{\to 0} = \frac{\gamma(0)}{4}(n_{i,+}(1 - n_{i-1,+}) + n_{i,-}(1 - n_{i-1,-}) + l_{i,+}^\dagger l_{i-1,+} l_{i-1,-}^\dagger l_{i,-} + l_{i,-}^\dagger l_{i-1,-} l_{i-1,+}^\dagger l_{i,+})$$

$$A_{i,a}^{\updownarrow 0} = \frac{\gamma(0)}{4}(n_{i-1,+}(1 - n_{i,+}) + n_{i-1,-}(1 - n_{i,-}) - l_{i-1,+}^\dagger l_{i,+} l_{i,-}^\dagger l_{i-1,-} - l_{i-1,-}^\dagger l_{i,-} l_{i,+}^\dagger l_{i-1,+})$$

$$A_{i,a}^{\to -} = \frac{\gamma(2)}{4}n_{i-1,-}(1 - n_{i,+}), \qquad A_{i,a}^{\updownarrow -} = \frac{\gamma(2)}{4}n_{i,+}(1 - n_{i-1,-})$$

Note that $A_{i,b}^{\leftrightarrow} = A_{i+1,a}^{\leftrightarrow}$. The jumps can only lower the system down an energy level if the hole and the excitation are spatially next to one another. Until that time, the system moves both excitations and holes by at most one site with frequency proportional to the magnitude of $\tilde{H}$. The effective Hamiltonian $\tilde{H}$ however is itself not fully diagonal in this basis, which was also encountered in the construction of the random walk for the engineered model. Thus, both protocols share the same locality constraints and structure of the effective Hamiltonian.



\end{document}